\documentclass[twocolumn]{aastex63}
\usepackage{float,graphicx,amsmath,multirow}
\usepackage{color}
\usepackage[version=3]{mhchem}
\usepackage{booktabs}

\begin{document}

\title{Searches for Interstellar HCCSH and \ce{H2CCS}}
\author{Brett A. McGuire}
\altaffiliation{B.A.M. is a Hubble Fellow of the National Radio Astronomy\\ Observatory.}
\affiliation{National Radio Astronomy Observatory, Charlottesville, VA 22903, USA}
\affiliation{Center for Astrophysics | Harvard \& Smithsonian, Cambridge, MA 02138, USA}
\author{Christopher N. Shingledecker}
\affiliation{Center for Astrochemical Studies, Max Planck Institute for Extraterrestrial Physics, Garching, Germany}
\affiliation{Institute for Theoretical Chemistry, University of Stuttgart, Stuttgart, Germany}
\author{Eric R. Willis}
\affiliation{Department of Chemistry, University of Virginia, Charlottesville, VA 22904, USA}
\author{Kin Long Kelvin Lee}
\affiliation{Center for Astrophysics | Harvard \& Smithsonian, Cambridge, MA 02138, USA}
\author{Marie-Aline Martin-Drumel}
\affiliation{Institut des Sciences Mol\'{e}culaires d'Orsay, CNRS, Univ Paris Sud, Universit\'{e} Paris-Saclay, Orsay, France}
\author{Geoffrey A. Blake}
\affiliation{Division of Planetary Sciences and Division of Chemistry and Chemical Engineering, California Institute of Technology, Pasadena, CA 91125, USA}
\author{Crystal L. Brogan}
\affiliation{National Radio Astronomy Observatory, Charlottesville, VA 22903, USA}
\author{Andrew M. Burkhardt}
\affiliation{Center for Astrophysics | Harvard \& Smithsonian, Cambridge, MA 02138, USA}
\author{Paola Caselli}
\affiliation{Center for Astrochemical Studies, Max Planck Intitute for Extraterrestrial Physics, Garching, Germany}
\author{Ko-Ju Chuang}
\affiliation{Sackler Laboratory for Astrophysics, Leiden Observatory, Leiden University, PO Box 9513, 2300 RA Leiden, The Netherlands}
\author{Samer El-Abd}
\affiliation{Department of Astronomy, University of Virginia, Charlottesville, VA 22904, USA}
\author{Todd R. Hunter}
\affiliation{National Radio Astronomy Observatory, Charlottesville, VA 22903, USA}
\author{Sergio Ioppolo}
\affiliation{School of Electronic Engineering and Computer Science, Queen Mary University of London, Mile End Road, London E1 4NS, UK}
\author{Harold Linnartz}
\affiliation{Sackler Laboratory for Astrophysics, Leiden Observatory, Leiden University, PO Box 9513, 2300 RA Leiden, The Netherlands}
\author{Anthony J. Remijan}
\affiliation{National Radio Astronomy Observatory, Charlottesville, VA 22903, USA}
\author{Ci Xue}
\affiliation{Department of Chemistry, University of Virginia, Charlottesville, VA 22904, USA}
\author{Michael C. McCarthy}
\affiliation{Center for Astrophysics | Harvard \& Smithsonian, Cambridge, MA 02138, USA}

\begin{abstract}

A long standing problem in astrochemistry is the inability of many current models to account for missing sulfur content.  Many relatively simple species that may be good candidates to sequester sulfur have not been measured experimentally at the high spectral resolution necessary to enable radioastronomical identification. On the basis of new laboratory data, we report searches for the rotational lines in the microwave, millimeter, and sub-millimeter regions of the sulfur-containing hydrocarbon HCCSH.   This simple species would appear to be a promising candidate for detection in space owing to the large  dipole moment along its $b$-inertial axis, and because the bimolecular reaction between two highly abundant astronomical fragments (CCH and SH radicals) may be rapid.  An inspection of multiple line surveys from the centimeter to the far-infrared toward a range of sources from dark clouds to high-mass star-forming regions, however, resulted in non-detections.  An analogous search for the lowest-energy isomer, \ce{H2CCS}, is presented for comparison, and also resulted in non-detections. Typical upper limits on the abundance of both species relative to hydrogen are $10^{-9}$--$10^{-10}$.  We thus conclude that neither isomer is a major reservoir of interstellar sulfur in the range of environments studied.  Both species may still be viable candidates for detection in other environments or at higher frequencies, providing laboratory frequencies are available. 

\end{abstract}

\section{Introduction}
\label{intro}

Despite being one of the most abundant elements in the interstellar medium (ISM), the sulfur content in cold molecular clouds can only account for $\sim$0.1\% of that seen in warm, diffuse clouds \citep{Tieftrunk:1994rk,Ruffle:1999fla}, which have  values comparable to the solar abundance \citep{Bilalbegovic:2015dr}.  To account for this ``missing" sulfur, many hypotheses have been put forward regarding hidden sinks and reservoirs, including ices, dusty grains, and unknown molecular species \citep{MartinDomenech:2016je}.   

The search for condensed-phase sulfur has largely been unsuccessful to date.  Simple molecules (such as OCS) sequestered in interstellar ice grains, for example, can only  account \citep{Boogert:1997ys,MartinDomenech:2016je} for a small amount of the total cosmic abundance ($\sim$4\%). In cometary ices, the principle sulfur-bearing species appears to be \ce{H2S}, with an abundance of $\sim$1.5\% relative to water \citep{Bockelee:2000ol}, whereas FeS appears to be a major reservoir in the cometary grains themselves \citep{Dai:2001iu}.  FeS grains have also been detected in protoplanetary disks, suggesting this reservoir may be widespread, and not unique to Solar System objects \citep{Keller:2002db}.  Several recent observational and modeling studies hypothesize  \ce{H2S} may be a substantial sulfur reservoir in interstellar ices as well \citep{Holdship:2016js,Vidal:2017gwa}. While \ce{H2S} is a well-known interstellar species in the gas-phase \citep{Thaddeus:1972oh}, there has been no definitive condensed-phase observation of it outside the Solar System \citep{Boogert:2015fx}.  Recent modeling work by \citet{Laas:2019kb} has shown that the sulfur depletion in cold clouds can be reproduced without the need for a significant build-up of \ce{H2S} in ices, suggesting this species may play a less substantial role than previously assumed. Further investigations of condensed-phase sulfur species will undoubtedly be a strong science driver for the forthcoming James Webb Space Telescope. 

New gas-phase sulfur-bearing molecules remain one of the most promising explanations for the missing sulfur content, and one that is addressable by extant radio facilities, assuming the molecules possess a permanent dipole moment and the necessary laboratory work is both available and complete \citep{Cazzoli:2016ew}.  Of the more than 200 known interstellar and circumstellar molecules, only 23  contain at least one sulfur atom (Table~\ref{sulfurmols}).  Perhaps more striking, of the 94 molecules with five or more atoms, three contain sulfur, whereas 30 contain at least one oxygen atom \citep{McGuire:2018mc}.  In total, sulfur-containing species comprise $\sim$10\% of the known molecular inventory of any size, whereas a third of all molecules are oxygen-containing.  Despite the lower interstellar abundance of sulfur relative to oxygen \citep{Cameron:1970dw}, the large disparity between the number of S- and O-bearing species is striking, and may indicate detections of large, sulfur-bearing species in space is simply limited by the lack of precise laboratory rest frequencies.

\begin{table}[h]
    \centering
    \caption{Known interstellar and circumstellar molecules containing at least one sulfur atom and references to the first detection of those species.}
    \begin{tabular}{l l}
    \toprule
    \toprule
    Molecule        &   Reference   \\
    \midrule
    \ce{CS}         &   \citet{Penzias:1971kw}\\
    \ce{SO}         &   \citet{Gottlieb:1973md}\\
    \ce{SiS}        &   \citet{Morris:1975im}\\
    \ce{NS}         &   \citet{Gottlieb:1975ke,Kuiper:1975er}\\
    \ce{SO+}        &   \citet{Turner:1992cj}\\
    \ce{SH+}        &   \citet{Benz:2010ei}\\  
    \ce{SH}         &   \citet{Neufeld:2012gz}\\
    \ce{NS+}        &   \citet{Cernicharo:2018bv}\\
    \ce{OCS}        &   \citet{Jefferts:1971hy}\\
    \ce{H2S}        &   \citet{Thaddeus:1972oh}\\
    \ce{SO2}        &   \citet{Snyder:1975cr}\\
    \ce{HCS+}       &   \citet{Thaddeus:1981uy}\\
    \ce{C2S}        &   \citet{Saito:1987fa} \\  
    \ce{S2H}        &   \citet{Fuente:2017gd}\\
    \ce{HCS}        &   \citet{Agundez:2018kh}\\
    \ce{HSC}        &   \citet{Agundez:2018kh}\\  
    \ce{H2CS}       &   \citet{Sinclair:1973ft}\\
    \ce{HNCS}       &   \citet{Frerking:1979yd}\\
    \ce{C3S}        &   \citet{Yamamoto:1987jd}\\
    \ce{HSCN}       &   \citet{Halfen:2009it}\\
    \ce{CH3SH}      &   \citet{Linke:1979dc}\\
    \ce{C5S}        &   \citet{Bell:1993ir}\\  
    \ce{CH3CH2SH}   &   \citet{Kolesnikova:2014fb}\\
    \bottomrule
    \end{tabular}
    \label{sulfurmols}
\end{table}

Large sulfur-bearing species (by interstellar standards) are perhaps improbable as substantial reservoirs of gas-phase sulfur, given the generally lower abundance of large molecules relative to simpler species (see, e.g., \citealt{Belloche:2013eba}).  Detailed studies of their formation chemistry, however, can provide insights into the underlying abundances (and potential reservoirs) of as-yet-undetected simpler precursor molecules.  As an example, one of the most promising ways to probe formation pathways is through the study of isomeric families.  Comparisons of the  [\ce{H4},\ce{C2},\ce{O2}] isomers glycolaldehyde, methyl formate, and acetic acid, for instance, have been used in attempts to infer branching fractions in the UV photodissociation of methanol (\ce{CH3OH}), from which most of the key precursor species for the [\ce{H4},\ce{C2},\ce{O2}] family are expected to form \citep{Laas:2011yd}.

Here, we present an observational counterpart to the recent laboratory investigation of the [\ce{H2},\ce{C2},S] isomeric family of molecules.  The three lowest-energy isomers in this family are \ce{H2CCS} (thioketene), \ce{HCCSH} (ethynethiol), and $c$-\ce{H2C2S} (thiirene).  The microwave and millimeter-wave spectrum of \ce{H2CCS} has been previously reported by \citet{Winnewisser:1980ic}, and several of the co-authors of the present work recently reported the laboratory spectrum of \ce{HCCSH} from the microwave to sub-millimeter wavelengths \citep{Lee:2018uw}.  Efforts to measure the spectrum of $c$-\ce{H2C2S} are currently underway.  Here, we summarize the astronomical search for \ce{H2CCS} and \ce{HCCSH} in a number of interstellar sources spanning the evolutionary spectrum from dark clouds to high-mass star-forming regions (HMSFRs).

HCCSH would appear to be a promising candidate for astronomical detection and potential sink for sulfur for two reasons:  (1) this isomer might be formed directly and efficiently via a recombination reaction involving two well-known astronomical radicals, SH and CCH \citep{Yamada:2002gm}, and (2) HCCSH does not possess a simple, readily identifiable rotational spectrum, since the dipole moment along its $a$-inertial axis is nearly zero (0.13\,D), while that along the $b$-axis is significantly larger (0.80\,D).  As a consequence, unlike \ce{H2CCS} whose rotational spectrum is characterized by a series of lines separated in frequency by ratios of integers at low-frequency, the same lines of HCCSH are extremely faint; instead, its $b$-type lines in the millimeter- and sub-millimeter bands are much more intense [i.e., $(\mu_b/\mu_a)^2 \sim 35$].

\section{Energetics and Spectroscopy}
\label{spectroscopy}

Electronic structure calculations \citep{Lee:2018uw} predict that the most stable isomeric arrangement of [\ce{H2},\ce{C2},S] is \ce{H2CCS}, followed by  HCCSH roughly 56\,kJ/mol (${\sim}$6770\,K) higher in energy, with the three-membered heterocycle $c$-\ce{H2C2S} lying 132\,kJ/mol \newline(${\sim}$15,900\,K) above ground. Nevertheless,  HCCSH can be formed by an exothermic and barrierless reaction involving SH and CCH, via reaction~(\ref{r1};  \citealt{Lee:2018uw}):
\begin{equation}
    \ce{SH + CCH -> HCCSH}
    \tag{R1}
    \label{r1}
\end{equation}
This  may  result in preferential production in astronomical sources in which the two radicals are prominent, although in the gas-phase it is possible that the exothermicity may result in the dissociation of a non-trivial fraction of the product, necessitating grain-surface production.  The  closely related [\ce{H2},\ce{C3},O] isomers provide a dramatic illustration of this effect.  Although $l$-propadienone (\ce{H2CCCO}) has not been observed in space despite considerable efforts \citep{Loomis:2015jh}, isoenergetic propynal (HCCC(O)H) is prominent  \citep{Irvine:1988dw}. Further, the considerably less stable cyclopropenone ($c$-\ce{H2C3O}) has also been detected in space \citep{Hollis:2006ih}. Taken together, these findings highlight the importance of kinetic factors in molecule formation and destruction.  

Thioketene (\ce{H2CCS}) has a linear heavy-atom backbone with $C_{2v}$ symmetry and \textit{ortho-para} spin statistics.  For this reason, it only has $a$-type rotational lines, but its dipole moment is sizable, 1.01(3)\,D \citep{Winnewisser:1980ic}.  At low temperature, its rotational spectrum consists of relatively closely spaced triplets that are harmonically-related spaced by $B+C$, or about 11.2\,GHz.  Laboratory measurements provide rest frequencies up to 230\,GHz, which, given the rigidity of the molecule, can be extrapolated with reasonable confidence to better than 0.5\,km~s$^{-1}$ up to $\sim$450\,GHz.

Although HCCSH has the same heavy atom backbone as \ce{H2CCS}, owing to a different bond order along the chain, it is an asymmetric top with a bent backbone, but one still close to the prolate limit ($\kappa$~=~-0.999), as defined by $\kappa$, Ray's asymmetry parameter \citep{Ray:1932yd}. A purely prolate molecule (cigar shaped, and the most common in the ISM; \citealt{McGuire:2018mc}) has $\kappa=-1$, whereas a purely oblate molecule (disk shaped) has $\kappa$~=~1. In contrast to \ce{H2CCS}, HCCSH has a relatively small calculated dipole moment along its $a$-axis (0.13\,D), but that along the $b$-axis (0.80\,D) is significant  \citep{Lee:2018uw}.  Because the $A$ rotational constant is large, 291.4\,GHz,  the most intense rotational lines lie at millimeter wavelengths. The detailed spectroscopic analysis of its $a$- and $b$-type transitions up to 660\,GHz was presented in \citet{Lee:2018uw}. 

\section{Analysis}

Column densities were calculated assuming a single-excitation temperature model following the formalisms of \citet{Hollis:2004uh} using Eq.~\ref{hollis} below, with corrections due to optical depth as described in \citet{Turner:1991um}.

\begin{equation}
N_T = \frac{1}{2}\frac{3k}{8\pi^3}\sqrt{\frac{\pi}{\ln 2}}\frac{Qe^{E_u/T_{\rm{ex}}}\Delta T_b\Delta V}{B\nu S_{ij}\mu^2}\frac{1}{1-\frac{e^{h\nu/kT_{\rm{ex}}} - 1}{e^{h\nu/kT_{\rm{bg}}}-1}}
\label{hollis}
\end{equation}

Here, $N_T$ is the total column density (cm$^{-2}$), $Q$ is the partition function, $T_{ex}$ is the excitation temperature (K), $E_u$ is the upper state energy (K), $\Delta T_b$ is the peak intensity (K), $\Delta V$ is the full-width at half-maximum of the line (km~s$^{-1}$), $B$ is the beam filling factor, $\nu$ is the frequency (Hz), $S_{ij}$ is the intrinsic quantum mechanical line strength, $\mu$ is the permanent dipole moment (Debye\footnote{Care must be taken to convert this unit for compatibility with the rest of the parameters.}), and $T_{bg}$ is the background continuum temperature (K).  For the upper-limits presented here, a simulated spectrum was generated using the parameters described for each observation below.  The strongest predicted line in the observational spectrum was then used to calculate the $1\sigma$ upper limit, taking the rms noise value at that point as the value of the brightness temperature $\Delta T_b$.

For HCSSH and \ce{H2CCS}, we consider both the rotational partition function and the contribution from low-lying vibrational states. The total partition function is calculated according to Eq.~\ref{q_tot},
\begin{equation}
    Q = Q_{rot} \times Q_{vib} \times Q_{elec}
    \label{q_tot}
\end{equation}
where $Q_{rot}$, $Q_{vib}$, and $Q_{elec}$ are the rotational, vibrational, and electronic components, respectively.  Under interstellar conditions, we assume $Q_{elec} = 1$.
The rotational partition function was calculated via a direct summation of states as described by \citet{Gordy:1984uy} and Eq.~\ref{q_rot},
\begin{equation}
\label{q_rot}
Q_r = \sum_{i=0}^{\infty}(2J+1)e^{{-E_i}/kT_{ex}}
\end{equation}
where $E_i$ is the energy of the $i^{th}$ rotational state.  The vibrational correction was calculated according to Eq.~\ref{q_vib},
\begin{equation}
    Q_{\rm{vib}}=\prod\limits_{\substack{i=1}}^{3N-6} \frac{1}{1-e^{-E_{v,i}/kT_{\rm{ex}}}}
    \label{q_vib}
\end{equation}
where $E_{v,i}$ is the energy of the $i^{th}$ excited vibrational state.  Here, we consider only the lowest five vibrational states as those higher in energy make a negligible contribution to $Q_{\rm{vib}}$.  

The vibrational (harmonic) energies were calculated at the CCSD(T)/cc-pVQZ level of theory.  For HCCSH, the lowest five vibrational states lie at 356, 416, 727, 848, and 942\,cm$^{-1}$ above ground, while for \ce{H2CCS}, they fall at 295, 352, 578, 696, and 718\,cm$^{-1}$.  The rotational, vibrational, and total partition functions used for the column density calculations at each temperature are listed in Tables~\ref{hccsh_ulims}~\&~\ref{h2ccs_ulims} along with other molecular parameters required for the calculations.

\section{Observations and Analysis}
\label{results}

\begin{figure*}[t!]
\includegraphics[width=\textwidth]{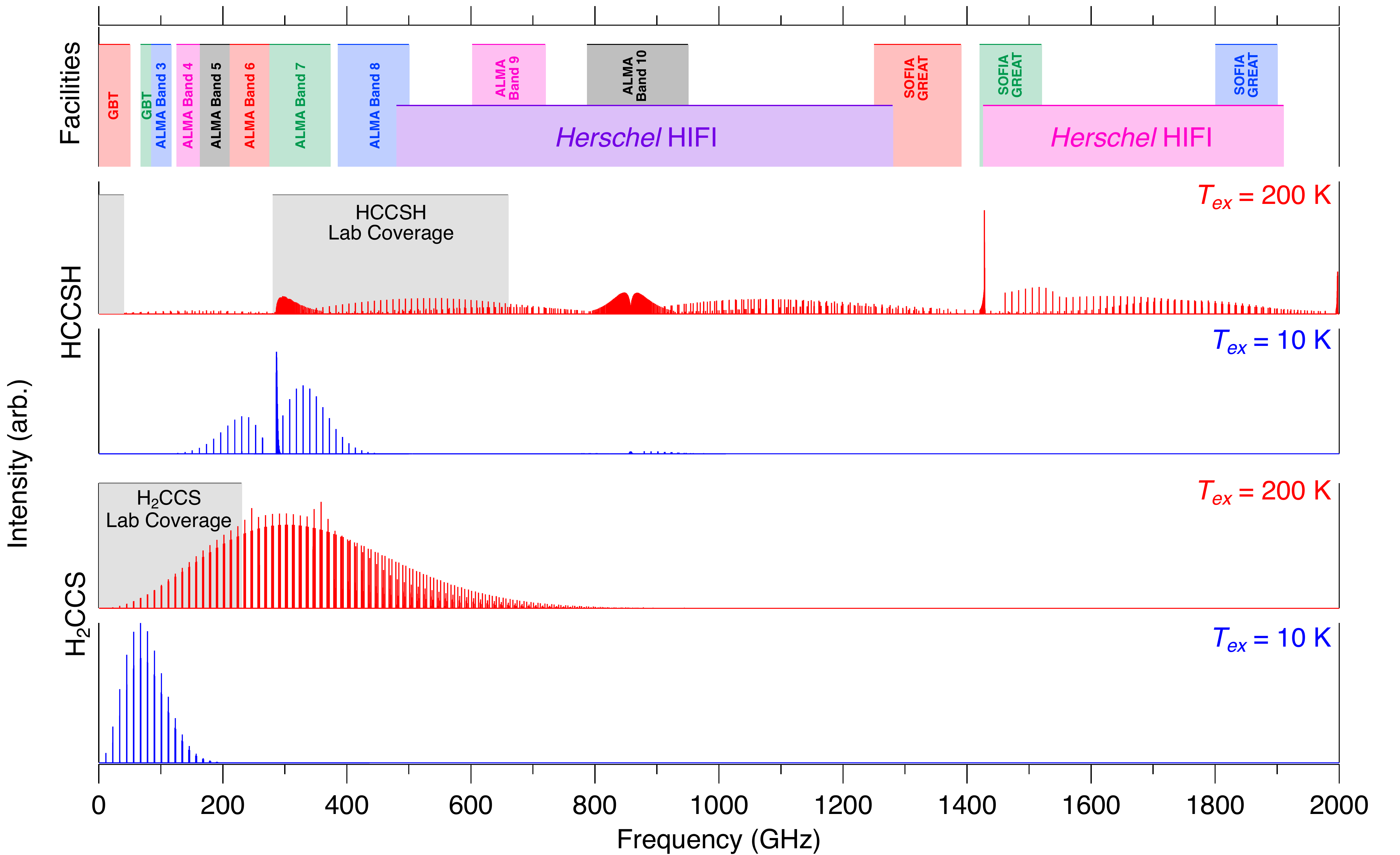}
\caption{Simulations of HCCSH and \ce{H2CCS} up to 2\,THz and 1 THz, respectively, at $T_{ex}$~=~10\,K (blue) and 200\,K (red).  The top panel of the figure shows the frequency coverage available for the GBT, ALMA, SOFIA, and in archival \emph{Herschel} observations.  The shaded gray regions show the extent of the frequency range that has been observed in the laboratory for these species.}
\label{coverage}
\end{figure*}

The simulated spectrum of HCCSH, with arbitrary abundance, at three different temperatures appropriate to interstellar environments (10, 80, and 200\,K) is shown in Fig.~\ref{coverage}, along with the associated observing bands of the most sensitive telescope facilities in those frequency ranges: the Robert C. Byrd 100 m Green Bank Telescope (GBT), the Atacama Large Millimeter/sub-millimeter Array (ALMA), and the GREAT instrument aboard Stratospheric Observatory for Infrared Astronomy (SOFIA). Due to its geometry, the most intense transitions fall in ALMA Bands 7 and 10 and the 1.4\,THz window of the SOFIA GREAT receiver , as well as archival coverage of the \emph{Herschel} HIFI instrument.  Here, we examined observations from ALMA, \emph{Herschel}, and the GBT targeting the HMSFRs NGC 6334I, Sgr B2(N), and Orion-KL.  We also searched the line survey data toward sources from the publicly available Astrochemical Surveys at IRAM (ASAI) Large Project conducted with the IRAM 30-m telescope.\footnote{\url{http://www.oan.es/asai/}} The exact physical parameters assumed for each of the sources examined here are given in Tables~\ref{hccsh_source_params} and \ref{h2ccs_source_params}.

\subsection{NGC 6334I}
\label{6334}

NGC 6334I is a nearby HMSFR (1.3~kpc by maser parallax; \citealt{Chibueze:2014iv}) with a number of embedded protostars \citep{Hunter:2006th,Brogan:2016cy,Hunter:2017th}, young, active, and variable outflows and molecular masers \citep{Hunter:2018gz,Brogan:2018wb,McGuire:2018bz}, and a rich molecular inventory \citep{McGuire:2017gy}.  The spectra examined here toward NGC 6334I were obtained with ALMA project codes \#2015.A.00022.T and \#2017.1.00717.S.   The spectra were extracted from (J2000) $\alpha$~=~17:20:53.374, $\delta$=-35:46:58.34.  This location is nearby the MM1 embedded protostar(s), but sufficiently far from the continuum peak to minimize absorption of molecular lines.

The complete observing parameters for the Band 10 data are provided in \citet{McGuire:2018bz}. The Band 7 observing parameters are provided in \citet{McGuire:2017gy}, with one exception.  To ensure a consistent dataset, the angular resolution of the Band 7 data, originally 0.25$^{\prime\prime}\times$0.19$^{\prime\prime}$ was degraded to match that of the Band 10 data: 0.26$^{\prime\prime}\times$0.26$^{\prime\prime}$.  The values of $T_{ex}$, $v_{lsr}$, $\Delta V$, and $T_{bg}$ for both HCCSH and \ce{H2CCS} were assumed to match those described in \citet{McGuire:2018bz} that were found to well-reproduce the complex molecular spectra observed across both Band 7 and 10.

\subsection{Orion-KL}
\label{orikl}

At a distance of 414$\pm$7~pc \citep{Menten:2007ew}, Orion-KL is perhaps the closest well-studied molecularly rich HMSFR.  Indeed, six sulfur-bearing molecules were detected first  toward Orion: CS, SO, \ce{H2S}, \ce{SO2}, \ce{HCS+}, and \ce{CH3CH2SH} (\citealt{McGuire:2018mc} and refs. therein).  Like NGC 6334I, it displays a complex physical structure, with a hot core, both compact and extended molecular ridges, and outflows (see Fig.~7 of \citealt{Crockett:2014er}).  The spectra examined here toward Orion KL around 860\,GHz are from the \emph{Herschel} Observations of EXtraordinary Sources (HEXOS) key project using the Heterodyne Instrument for the Far Infrared (HIFI) instrument on the \emph{Herschel Space Observatory}.  The details of the observations are presented in \citet{Crockett:2014er}.  The background continuum temperature was obtained by a linear fit to the spectra prior to continuum subtraction.

For HCCSH, the values of $T_{ex}$, $v_{lsr}$, $\Delta V$, and $\theta_s$ were assumed to match those of \ce{HN^{13}CO} from the analysis of \citet{Crockett:2014er}, since this molecule is the most structurally similar one (a three heavy atom backbone with an off-axis hydrogen) to HCCSH that was not optically thick.   The strongest transitions of HCCSH fall around 850\,GHz (see Fig. \ref{coverage}), but these are both outside of the range of the laboratory measurements ($\nu_{max}$ = 660\,GHz), and are of a type ($J^{\prime}_{2,K_c} - J^{\prime\prime}_{1,K_c}$) not fit by the laboratory measurements.  The extrapolated frequencies are likely accurate enough to confirm a non-detection, but because of these uncertainties, we also report an upper limit derived from transitions that, while weaker, fall within the range of the laboratory measurements. 

For \ce{H2CCS}, we adopted the parameters from \citet{Crockett:2014er} for its oxygen analog \ce{H2CCO}.  At the assumed excitation temperature of $T_{ex}$~=~100~K, the strongest lines of \ce{H2CCS} fall around 280~GHz, 200~GHz lower than the coverage of the HEXOS observations.  A further reduction on the intensity of its transitions in the lowest HEXOS band occurs because of a small source size ($\theta_s$~=~10$^{\prime\prime}$).  These factors, combined with the need to extrapolate the molecular fit substantially above the measured laboratory data, make the derived upper limit for \ce{H2CCS} in this source relatively uncertain.

\subsection{Sgr B2(N)}
\label{sgrb2}

Sgr B2(N) is the premier hunting ground for new molecular detections in HMSFRs.  Located at a distance of 8.3~kpc \citep{Reid:2014km}, this complex contains a number of embedded molecular cores separated by of order a few arcseconds (see, e.g., \citealt{Bonfand:2017eo}).  Molecules are typically detected in one of two regimes: either warm and compact, in the hot cores of the region (e.g., \citealt{Belloche:2013eba}), or cold, diffuse, and in absorption (normally only at low frequencies), in an extended molecular shell around the region (e.g., \citealt{McGuire:2016ba}).  We examined three datasets for Sgr B2(N) covering a range of these conditions: the \emph{Herschel} HEXOS survey at sub-millimeter wavelengths, the IRAM 30\,m survey at millimeter wavelengths, and the GBT PRIMOS survey at centimeter wavelengths.

\subsubsection{\emph{Herschel} Data} 

The spectra used here toward Sgr B2(N) around 860~GHz are also from the HEXOS key project using the HIFI instrument on the \emph{Herschel Space Observatory}.  The details of the observations are presented in \citet{Neill:2014cb}.  The background continuum temperature was obtained by a linear fit to the spectra prior to background subtraction. As with the HEXOS observations of Orion-KL, the values of $T_{ex}$, $v_{lsr}$, $\Delta V$, and $\theta_s$ were assumed to match those of isocyanic acid, HNCO for HCCSH, and upper limits were derived from both the strongest predicted lines, and from lines within the range of the laboratory measurements.  Although substantial absorption is seen from HNCO at colder temperatures with an extended source size distribution, the warm, compact component of \ce{HN^{12}CO} in emission is optically thin.  Unlike in Orion-KL, \ce{H2CCO} is not detected in Sgr B2(N), and thus we also use these same parameters derived from HNCO for \ce{H2CCS}.

\subsubsection{IRAM 30\,m Data} 

At even modest excitation temperatures, the transitions of \ce{HCCSH} arising in the millimeter and centimeter are too weak to provide a meaningful comparison to the limits established by the \emph{Herschel} data, however, strong transitions of \ce{H2CCS} are still present.   A dataset at 3~mm from the IRAM 30-m telescope is available from the work of \citet{Belloche:2013eba}, and we assume for \ce{H2CCS} the physical parameters derived for \ce{H2CCO} from that work for the upper limit calculations.

\subsubsection{GBT PRIMOS Data}

Finally, if the \ce{H2CCS} is particularly cold, a number of low-$J$ transitions will show bright absorption against background continuum at centimeter-wavelengths, whereas no reasonably intense signal, either in absorption or emission, can be expected for HCCSH because its transitions at these frequencies are very weak.  This frequency range falls within the coverage of the Prebiotic Interstellar Molecular Survey (PRIMOS) project observations of Sgr B2(N) using the 100-m Robert C. Byrd Green Bank Telescope.  Observational details and data reduction procedures are outlined in \citet{Neill:2012fr}.   In the case of cold molecules observed in absorption toward Sgr B2(N), the bright background continuum against which this absorption is seen is non-thermal \citep{Hollis:2007ww}, and has a source size of $\theta_s$~$\sim$20$^{\prime\prime}$ \citep{Mehringer:1993dd}.  These molecules are typically well described by a single, sub-thermal $T_{ex}$~$\sim$5~K \citep{McGuire:2016ba}.  Here, we have adopted the parameters that well-model the observed absorption signal from acetone (\ce{CH3C(O)CH3}), as well as the detailed modeling of the background continuum, source size, and beam size effects, as described in the Supplementary Material of \citet{McGuire:2016ba}.

\subsection{ASAI Sources}

Observational details of the ASAI sources are presented in \citet{Lefloch:2018fk}; these cover a range of Solar-type protostellar sources from dark clouds to Class 0/1 protostars, including shocked outflows.  Because the chemical inventories of these sources are quite varied, for the purposes of this work, we have adopted source parameters and excitation conditions for HCCSH and \ce{H2CCS} representative of complex molecules previously seen in these sources, as gathered from the literature.

\section{Results and Discussion}
\label{discussion}

\begin{table*}[!htb]
    \scriptsize
    \centering
    \caption{Upper limits to HCCSH and the line parameters used to calculate them in each of the sets of observations.}
    \begin{tabular}{l r c c c r c c c c }
    \toprule
    \toprule
    Source              &   \multicolumn{1}{c}{Frequency$^a$}   &   Transition          &   $E_u$   &   $S_{ij}\mu^{2}$ & \multicolumn{1}{c}{$Q$ ($Q_{rot}$, $Q_{vib}$)$^c$}    & $N_T$                     & $N(\ce{H2})$                  & $X_{H_2}$                 &    Refs.           \\
                        &   \multicolumn{1}{c}{(MHz)}           &   ($J_{K_a,K_c}^{\prime} - J_{K_a,K_c}^{\prime\prime}$) & (K)   &   (Debye$^2$) &                                 & (cm$^{-2}$)               & (cm$^{-2}$)                   &                           &    $N(\ce{H2})$  \\
    \midrule
    NGC 6334I           &   293352.8        &   $16_{1,15} - 16_{0,16}$                     &   85.8    &   10.4            & 3894 (2824, 1.38)                                     & $\leq 1.4\times10^{16}$   & --                            & --                        &  --                  \\
    Sgr B2(N)           &   618565.8        &   $33_{1,33} - 32_{0,32}$                     &   307.9   &   11.4            & 35808 (8441, 4.24)                                    & $\leq 3.9\times10^{17}$   & $1\times10^{24}$              & $\leq 4\times10^{-7}$     &  1                  \\
    Sgr B2(N)           &   844982.2$^b$    &   $21_{2,19} - 21_{1,20}$                     &   176.7   &   7.1             & 35808 (8441, 4.24)                                    & $\leq 1.4\times10^{17}$   & $1\times10^{24}$              & $\leq 1\times10^{-7}$     &  1                  \\
    Orion-KL            &   618565.8        &   $33_{1,33} - 32_{0,32}$                     &   307.9   &   11.4            & 12664 (5442, 2.32)                                    & $\leq 3.1\times10^{15}$   & $3.9\times10^{23}$            & $\leq 8\times10^{-9}$     &  2                  \\
    Orion-KL            &   850042.2$^b$    &   $16_{2,14} - 16_{1,15}$                     &   126.6   &   5.3             & 12664 (5442, 2.32)                                    & $\leq 3.3\times10^{15}$   & $3.9\times10^{23}$            & $\leq 8\times10^{-9}$     &  2                  \\
    &\vspace{-0.5em}    \\
    Barnard 1           &   241629.5        &   $3_{1,3} - 4_{0,4}$                         &   16.9    &   1.0             & 57 (57, 1.00)                                         & $\leq 2.3\times10^{12}$   & $1.5\times10^{23}$            & $\leq 2\times10^{-11}$    &   3                 \\
    IRAS 4A             &   230425.1        &   $4_{1,4} - 5_{0,5}$                         &   19.0    &   1.3             & 173 (173, 1.00)                                       & $\leq 1.3\times10^{13}$   & $3.7\times10^{23}$            & $\leq 4\times10^{-11}$    &   3                 \\
    L1157B1             &   288898.1        &   $10_{1,9} - 10_{0,10}$                      &   42.9    &   6.7             & 855 (837, 1.02)                                       & $\leq 5.3\times10^{12}$   & $1\times10^{21}$              & $\leq 5\times10^{-9}$     &   3                 \\
    L1157mm             &   207853.8        &   $6_{1,6} - 7_{0,7}$                         &   24.7    &   1.9             & 855 (837, 1.02)                                       & $\leq 7.7\times10^{12}$   & $6\times10^{21}$              & $\leq 1\times10^{-9}$    &   3                 \\
    L1448R2             &   207853.8        &   $6_{1,6} - 7_{0,7}$                         &   24.7    &   1.9             & 855 (837, 1.02)                                       & $\leq 2.1\times10^{13}$   & $3.5\times10^{23}$            & $\leq 6\times10^{-11}$    &   4                 \\
    L1527               &   162066.9        &   $10_{1,10} - 11_{0,11}$                     &   42.6    &   3.2             & 75 (75, 1.00)                                         & $\leq 3.1\times10^{12}$   & $2.8\times10^{22}$            & $\leq 1\times10^{-10}$    &   4                 \\
    L1544               &   87882.4         &   $8_{0,8} - 7_{0,7}$                         &   19.0    &   0.1             & 57 (57, 1.00)                                         & $\leq 1.6\times10^{13}$   & $5\times10^{21}$              & $\leq 3\times10^{-9}$    &   5                 \\
    SVS13A              &   207853.8        &   $6_{1,6} - 7_{0,7}$                         &   24.7    &   1.9             & 149 (149, 1.00)                                       & $\leq 9.0\times10^{15}$   & $3\times10^{24}$              & $\leq 3\times10^{-9}$     &   6                 \\
    TMC1                &   150487.4        &   $11_{1,11} - 12_{0,12}$                     &   48.3    &   3.6             & 34 (34, 1.00)                                         & $\leq 2.9\times10^{13}$   & $1\times10^{22}$              & $\leq 3\times10^{-9}$    &   3                 \\  
    \bottomrule
    \end{tabular}
    \begin{minipage}{.94\textwidth}
		\footnotesize
		$^a$Typical experimental accuracy of the mm-wave measurements was $\sim$50\,kHz.\\
		$^b$Extrapolated beyond the upper range (660\,GHz) of the laboratory measurements.\\
		$^c$Calculated at the excitation temperature assumed for the source. See Table~\ref{hccsh_source_params}.\\
		\textbf{References -- } [1] \citealt{Lis:1990tw} [2] \citealt{Crockett:2014er} [3] \citealt{Cernicharo:2018bv} [4] \citealt{Jorgensen:2002ci}  [5] \citealt{Vastel:2014ev} [6] \citealt{Chen:2009jj}    
	\end{minipage}    
    \label{hccsh_ulims}
\end{table*}

\begin{table*}[!htb]
    \scriptsize
    \centering
    \caption{Upper limits to \ce{H2CCS} and the line parameters used to calculate them in each of the sets of observations.}
    \begin{tabular}{l r c c c r c c c c }
    \toprule
    \toprule
    Source              &   \multicolumn{1}{c}{Frequency$^a$}   &   Transition          &   $E_u$   &   $S_{ij}\mu^{2}$ & \multicolumn{1}{c}{$Q$ ($Q_{rot}$, $Q_{vib}$)$^f$}            & $N_T$                     & $N(\ce{H2})$                  & $X_{H_2}$                 &    Refs.           \\
                        &   \multicolumn{1}{c}{(MHz)}           &   ($J_{K_a,K_c}^{\prime} - J_{K_a,K_c}^{\prime\prime}$) & (K)   &   (Debye$^2$) &                                 & (cm$^{-2}$)               & (cm$^{-2}$)                   &                           &    $N(\ce{H2})$  \\
    \midrule
    NGC 6334I           &   292685.4$^b$    &   $26_{1,25} - 25_{1,24}$                     &   203.2   &   78.2            & 9051 (5456, 1.66)                                     & $\leq 4.7\times10^{15}$   & --                            & --                        &  --                  \\
    Sgr B2(N)$^c$       &   22407.9         &   $2_{0,2} - 1_{0,1}$                         &   1.6     &   2.0             & 31 (31, 1.00)                                         & $\leq 6.4\times10^{12}$   & $1\times10^{24}$              & $\leq 6\times10^{-12}$    &  1                  \\
    Sgr B2(N)$^d$       &   100828.5        &   $9_{0,9} - 8_{0,8}$                         &   24.2    &   9.0             & 12106 (6395, 1.89)                                    & $\leq 2.9\times10^{16}$   & $1\times10^{24}$              & $\leq 3\times10^{-8}$     &  1                  \\
    Sgr B2(N)$^e$       & 494982.4$^b$      &   $44_{1,43} - 43_{1,42}$                     &   548.3   &   132.5           & 121982 (16448, 7.42)                                  & $\leq 2.7\times10^{17}$   & $1\times10^{24}$              & $\leq 3\times10^{-7}$     &  1                  \\
    Orion-KL            &   494982.4$^b$    &   $44_{1,43} - 43_{1,42}$                     &   548.3   &   132.5           & 4410 (3466, 1.27)                                     & $\leq 3.2\times10^{15}$   & $3.9\times10^{23}$            & $\leq 8\times10^{-9}$     &  2                  \\
    &\vspace{-0.5em}    \\
    Barnard 1           &   89626.7         &   $8_{0,8} - 7_{0,7}$                         &   19.4    &   8.2             & 104 (104, 1.00)                                       & $\leq 5.0\times10^{11}$   & $1.5\times10^{23}$            & $\leq 3\times10^{-12}$    &   3                 \\
    IRAS 4A             &   89626.7         &   $8_{0,8} - 7_{0,7}$                         &   19.4    &   8.2             & 334 (334, 1.00)                                       & $\leq 2.5\times10^{12}$   & $3.7\times10^{23}$            & $\leq 7\times10^{-12}$    &   3                 \\
    L1157B1             &   213921.0        &   $19_{1,18} - 18_{1,17}$                     &   116.2   &   58.0            & 1725 (1648, 1.05)                                     & $\leq 3.1\times10^{12}$   & $1\times10^{21}$              & $\leq 3\times10^{-9}$     &   3                 \\
    L1157mm             &   211738.3        &   $19_{1,19} - 18_{1,18}$                     &   115.1   &   58.0            & 1725 (1648, 1.05)                                     & $\leq 2.4\times10^{12}$   & $6\times10^{21}$              & $\leq 4\times10^{-10}$    &   3                 \\
    L1448R2             &   100311.4        &   $9_{1,9} - 8_{1,8}$                         &   37.6    &   27.2            & 1725 (1648, 1.05)                                     & $\leq 4.2\times10^{12}$   & $3.5\times10^{23}$            & $\leq 1\times10^{-11}$    &   4                 \\
    L1527               &   78825.9         &   $7_{1,6} - 6_{1,5}$                         &   28.6    &   21.0            & 139 (139, 1.00)                                       & $\leq 3.1\times10^{11}$   & $2.8\times10^{22}$            & $\leq 1\times10^{-11}$    &   4                 \\
    L1544               &   89626.7         &   $8_{0,8} - 7_{0,7}$                         &   19.4    &   8.2             & 104 (104, 1.00)                                       & $\leq 2.2\times10^{11}$   & $5\times10^{21}$              & $\leq 4\times10^{-11}$    &   5                 \\
    SVS13A              &   100828.5        &   $9_{0,9} - 8_{0,8}$                         &   24.2    &   9.2             & 286 (286, 1.00)                                       & $\leq 9.1\times10^{15}$   & $3\times10^{24}$              & $\leq 3\times10^{-9}$     &   6                 \\
    TMC1                &   134430.3        &   $12_{0,12} - 11_{0,11}$                     &   41.9    &   12.2            & 57 (57, 1.00)                                         & $\leq 5.5\times10^{12}$   & $1\times10^{22}$              & $\leq 6\times10^{-10}$    &   3                 \\  
    \bottomrule
    \end{tabular}
    \begin{minipage}{.94\textwidth}
		\footnotesize
		$^a$Within the range of the measurements (60--230\,GHz), \citealt{Winnewisser:1980ic} claim a typical accuracy of $\sim$16.5\,kHz.\\
		$^b$Extrapolated beyond the upper range (230\,GHz) of the laboratory measurements.\\
		$^c$GBT (PRIMOS) Observations.\\
		$^d$IRAM 30-m Observations.\\
		$^e$\emph{Herschel} HIFI Observations.\\
		$^f$Calculated at the excitation temperature assumed for the source. See Table~\ref{h2ccs_source_params}.\\
		\textbf{References -- } [1] \citealt{Lis:1990tw} [2] \citealt{Crockett:2014er} [3] \citealt{Cernicharo:2018bv} [4] \citealt{Jorgensen:2002ci}  [5] \citealt{Vastel:2014ev} [6] \citealt{Chen:2009jj}    
	\end{minipage}    
    \label{h2ccs_ulims}
\end{table*}

In addition to the formation of HCCSH via reaction \eqref{r1}, as proposed by \citet{Lee:2018uw}, several additional reaction pathways would appear at least qualitatively plausible. For example, both quantum chemical \citep{Ochsenfeld:1999fz} and laboratory work \citep{Galland:2001ei} have shown that HCS/HSC can be readily produced by a reaction involving atomic carbon:
\begin{equation}
    \ce{H2S + C -> HCS/HSC + H}.
    \tag{R2}
    \label{r2}
\end{equation}
It is not unreasonable, therefore, to speculate that HCS/HSC could further react with atomic carbon, followed by hydrogenation, to yield both HCCSH and \ce{H2CCS} on grain surfaces.  Indeed, successive hydrogenation reactions beginning from CS or \ce{C2S} could also yield HCS/HSC intermediates, or perhaps even HCCSH and \ce{H2CCS} directly (see, e.g., \citealt{Lamberts:2018eh}).  Further quantum chemical and laboratory work exploring the efficiency, branching ratios, and rates of these reactions would certainly help to shed light on the viability of these pathways.


It is also important to consider the potential destruction pathways of these species as well, and in this context it may be enlightening to compare the [\ce{H2},\ce{C2},S] species with a similar family of isomers having the formula [\ce{H2},\ce{C3},O]. As mentioned earlier, one rather longstanding astronomical mystery has been why the most stable of these isomers, propadienone (\ce{H2CCCO}) has thus far remained undetected in space \citep{Loomis:2015jh,Loison:2016ck}, despite detections of two higher-energy forms, propynal (\ce{HCCCHO}) and cyclopropenone ($c$-\ce{H2C3O}) \citep{Irvine:1988dw,Hollis:2006ih}, there. Recently, an ab initio study involving reactions between atomic hydrogen and propadienone and  propynal revealed unexpectedly that only the addition to propadienone, i.e.

\begin{equation}
    \ce{H + H2CCCO -> H2CCHCO,}
    \tag{R3}
    \label{r3}
\end{equation}

\noindent
was barrierless and exothermic \citep{Shingledecker:2019js}. Moreover, it was found that the radical formed via \eqref{r3} could again subsequently react with H to form propenal (\ce{CH2CHCHO}), a species which in fact has been observed by \citet{Hollis:2004oc} in Sgr B2(N), where the other isomers of [\ce{H2},\ce{C3},O] have been detected. Thus, reaction \eqref{r3} likely keeps the abundance of propadienone low, both on grains -- where it serves as a precursor to propenal -- and in the gas, where the association product likely dissociates. 

Similarly, given the high mobility of atomic hydrogen on grain surfaces -- particularly in warm environments -- it may be the case that \ce{H2CCS} and/or \ce{HCCSH} are efficiently destroyed by H. Intriguingly, the detection of the related saturated species \ce{CH3CH2SH} in Orion-KL  by \citet{Kolesnikova:2014fb} hints at the successive hydrogenation of a CCS backbone, with either of the linear [\ce{H2},\ce{C2},S] isomers potentially serving as precursors. Detailed calculations of these types of reactions could therefore reveal whether such kinetic effects might play a role in explaining the observational results described here.

Given that the structurally analogous \ce{H2CCC}, \ce{H2CCN}, and \ce{H2CCO} are all known interstellar species \citep{Cernicharo:1991iv,Irvine:1988ej,Turner:1977ig}, the non-detections of both \ce{HCCSH} and \ce{H2CCS} is somewhat surprising.  In a few cases, there are hints of emission at appropriate frequencies, but nevertheless, no signal was seen that could be even tentatively assigned with any confidence to either HCCSH or \ce{H2CCS}.  Upper limits to the column densities were established for each molecule in each source and are summarized in Tables~\ref{hccsh_ulims} and \ref{h2ccs_ulims}, along with the pertinent line parameters.   The observational spectra around each transition used to calculate the upper limits, as well as a simulation of the molecular spectra using those upper limits, are shown in Figures~\ref{hccsh_fig} and \ref{h2ccs_fig}.

Given the large number of non-detections presented here, we conclude that neither \ce{H2CCS} nor \ce{HCCSH} are substantial interstellar sulfur reservoirs.  These species, along with $c$-\ce{H2C2S}, however, remain reasonable candidates for interstellar detection in  sensitive observations, but probably in a source very rich in sulfur-bearing species.  We note that the upper limits reported here could be substantially improved if additional laboratory spectroscopy is performed.  As shown in Fig.~\ref{coverage}, the laboratory data for both HCCSH and \ce{H2CCS} fail to cover the strongest transitions of these species in warm environments.  In sources where line confusion is not an issue, the errors introduced by extrapolation would likely not preclude detection, but would instead require additional observing time owing to the need for wider frequency coverage.  In line-confused sources, however, such as NGC 6334I in ALMA Band 10, frequency extrapolation can not be made with confidence, making any purported detection highly tenuous.  Similarly, while the strong $b$-type branch of HCCSH near 1.4~THz is likely to be identifiable with SOFIA even if the frequencies are slightly uncertain, the search space would be substantially narrowed by laboratory measurements in this band.  Finally, a search for the higher-energy cyclic isomer, \emph{c}-\ce{H2C2S}, is currently impossible due to the lack of any laboratory spectra.

\section{Conclusions}
\label{conclusions}

A search for HCCSH and \ce{H2CCS} in a number of astronomical line surveys  based on newly-reported laboratory rest-frequencies obtained with a combination of microwave and sub-millimeter spectroscopy is presented.  Non-detections are reported in all  sources, suggesting that these molecules are not substantial reservoirs of sulfur, nor can they be readily used to infer sulfur chemistry.  A possible explanation for the absence of HCCSH is an analogous destruction pathway to \emph{l}-propadienone by barrierless, exothermic reaction with atomic hydrogen in the solid state.  The detection of the strongest transitions of HCCSH, which arise at THz frequencies, remains a possibility, and would be substantially aided by additional enabling laboratory work.

\acknowledgments

This manuscript makes use of the following ALMA data: ADS/JAO.ALMA\#2015.A.00022.T and \#2017.1.00717.S.  ALMA is a partnership of ESO (representing its member states), NSF (USA) and NINS (Japan), together with NRC (Canada) and NSC and ASIAA (Taiwan) and KASI (Republic of Korea), in cooperation with the Republic of Chile.  The Joint ALMA Observatory is operated by ESO, AUI/NRAO and NAOJ.  The National Radio Astronomy Observatory is a facility of the National Science Foundation operated under cooperative agreement by Associated Universities, Inc. The Green Bank Observatory is a facility of the National Science Foundation operated under cooperative agreement by Associated Universities, Inc. Support for B.A.M. was provided by NASA through Hubble Fellowship grant \#HST-HF2-51396 awarded by the Space Telescope Science Institute, which is operated by the Association of Universities for Research in Astronomy, Inc., for NASA, under contract NAS5-26555. C. N. S. thanks the Alexander von Humboldt Stiftung/Foundation for their generous support. M. C. M. acknowledges support from NASA grants NNX13AE59G and 80NSSC18K0396, and NSF grant AST--1615847. M.-A. M.-D. is thankful to the Programme National ``Physique et Chimie du Milieu Interstellaire'' (PCMI) of CNRS/INSU with INC/INP co-funded by CEA and CNES for support.

\appendix

\section{Source Parameters}
\label{source_params}

\renewcommand{\thefigure}{A\arabic{figure}}
\renewcommand{\thetable}{A\arabic{table}}
\renewcommand{\theequation}{A\arabic{equation}}
\setcounter{figure}{0}
\setcounter{table}{0}
\setcounter{equation}{0}

The physical parameters assumed for each source examined here are provided in Tables~\ref{hccsh_source_params} and \ref{h2ccs_source_params}.

\begin{table*}[h!]
    \scriptsize
    \centering
    \setlength{\tabcolsep}{1.5mm}
    \caption{Source parameters assumed for HCCSH in each of the sets of observations.}
    \begin{tabular}{l c c c c c c c c}
    \toprule
    \toprule
    Source              &   Telescope       & $\theta _s^a$         &  $T_{bg}$ & $\Delta V$    & $T_b^{\dagger}$        & $T_{ex}$  & Refs.    &   Notes    \\
                        &                   & ($^{\prime\prime}$)   &  (K)      & (km s$^{-1}$) & (mK)                   & (K)       &             \\
    \midrule
    NGC 6334I           &   ALMA            & --                    & 28.2      & 3.2           & 16.0$^b$                 & 135       & 1   \\
    Sgr B2(N)           &   \emph{Herschel} & 2.3                   & 7.1       & 8.0           & 81.8                   & 280       & 2        &   At 619\,GHz\\
    Sgr B2(N)           &   \emph{Herschel} & 2.3                   & 10.9      & 8.0           & 77.0                   & 280       & 2        &   At 845\,GHz\\
    Orion-KL            &   \emph{Herschel} & 10                    & 5.5       & 6.5           & 27.5                   & 209       & 3        &   At 619\,GHz\\
    Orion-KL            &   \emph{Herschel} & 10                    & 8.7       & 6.5           & 78.0                   & 209       & 3        &   At 850\,GHz\\
    &\vspace{-0.5em}    \\
    Barnard 1           &   IRAM            & --                    & 2.7       & 0.8           & 12.1                   & 10        & 4, 5  \\
    IRAS 4A             &   IRAM            & --                    & 2.7       & 5.0           & 10.3                    & 21        & 4, 6  \\
    L1157B1             &   IRAM            & --                    & 2.7       & 8.0           & 4.4                    & 60        & 7  \\
    L1157mm             &   IRAM            & --                    & 2.7       & 3.0           & 4.7                    & 60        & 7  \\
    L1448R2             &   IRAM            & --                    & 2.7       & 8.0           & 4.7                    & 60        & 8  \\
    L1527               &   IRAM            & --                    & 2.7       & 0.5           & 6.8                    & 12        & 8, 9  \\
    L1544               &   IRAM            & --                    & 2.7       & 0.5           & 2.9                    & 10        & 10, 11  \\
    SVS13A              &   IRAM            & 0.3                   & 2.7       & 3.0           & 7.8                    & 80        & 4, 6  \\
    TMC1                &   IRAM            & --                    & 2.7       & 0.3           & 7.6                    & 7         & 12, 13  \\
    \bottomrule
    \end{tabular}
    \begin{minipage}{\columnwidth}
		\scriptsize
		$^a$Except where noted, the source is assumed to fill the beam.\\
		$^b$For these interferometric observations, the intensity is given in mJy/beam rather than mK.\\
		$^{\dagger}$Taken either as the 1$\sigma$ RMS noise level at the location of the target line, or for line confusion limited spectra, the reported RMS noise of the observations.\\
		\textbf{References -- } [1] \citealt{McGuire:2018bz} [2] \citealt{Neill:2014cb} [3] \citealt{Crockett:2014er} [4] \citealt{Melosso:2018kl} [5] \citealt{Cernicharo:2018bv} [6] \citealt{Higuchi:2018dx} [7] \citealt{McGuire:2015bp} [8]\citealt{Jorgensen:2002ci} [9] \citet{Araki:2017kg} [10] \citealt{HilyBlant:2018ix} [11]  \citealt{Crapsi:2005kp} [12] \citealt{McGuire:2018it} [13] \citealt{Gratier:2016fj}
	\end{minipage}    
    \label{hccsh_source_params}
\end{table*}

\begin{table*}[h!]
    \scriptsize
    \centering
    \setlength{\tabcolsep}{1.5mm}
    \caption{Source parameters assumed for \ce{H2CCS} in each of the sets of observations.}
    \begin{tabular}{l c c c c c c c}
    \toprule
    \toprule
    Source              &   Telescope       & $\theta _s^a$         &  $T_{bg}$ & $\Delta V$    & $T_b^{\dagger}$        & $T_{ex}$  & Refs.     \\
                        &                   & ($^{\prime\prime}$)   &  (K)      & (km s$^{-1}$) & (mK)                   & (K)       &             \\
    \midrule
    NGC 6334I           &   ALMA            & --                    & 28.2      & 3.2           & 2.0$^b$                 & 135       & 1   \\
    Sgr B2(N)           &   GBT             & 20                    & 28.4      & 12.0          & -4.5                   & 5        & 14    \\
    Sgr B2(N)           &   IRAM            & 2.2                   & 5.2       & 7.0           & 12.0                    & 150       & 15    \\
    Sgr B2(N)           &   \emph{Herschel} & 2.3                   & 5.1       & 8.0           & 42.1                   & 280       & 2    \\
    Orion-KL            &   \emph{Herschel} & 10                    & 4.3       & 3.0           & 20.0                   & 100       & 3    \\
    &\vspace{-0.5em}    \\
    Barnard 1           &   IRAM            & --                    & 2.7       & 0.8           & 2.9                   & 10        & 4, 5  \\
    IRAS 4A             &   IRAM            & --                    & 2.7       & 5.0           & 2.4                    & 21        & 4, 6  \\
    L1157B1             &   IRAM            & --                    & 2.7       & 8.0           & 2.4                    & 60        & 7  \\
    L1157mm             &   IRAM            & --                    & 2.7       & 3.0           & 4.9                    & 60        & 7  \\
    L1448R2             &   IRAM            & --                    & 2.7       & 8.0           & 2.6                    & 60        & 8  \\
    L1527               &   IRAM            & --                    & 2.7       & 0.5           & 3.5                    & 12        & 8, 9  \\
    L1544               &   IRAM            & --                    & 2.7       & 0.5           & 2.0                    & 10        & 10, 11  \\
    SVS13A              &   IRAM            & 0.3                   & 2.7       & 3.0           & 2.2                    & 80        & 4, 6  \\
    TMC1                &   IRAM            & --                    & 2.7       & 0.3           & 6.6                   & 7         & 12, 13  \\
    \bottomrule
    \end{tabular}
    \begin{minipage}{\columnwidth}
		\scriptsize
		$^a$Except where noted, the source is assumed to fill the beam.\\
		$^b$For these interferometric observations, the intensity is given in mJy/beam rather than mK.\\
		$^{\dagger}$Taken either as the 1$\sigma$ RMS noise level at the location of the target line, or for line confusion limited spectra, the reported RMS noise of the observations.\\
		\textbf{References -- } [1] \citealt{McGuire:2018bz} [2] \citealt{Neill:2014cb} [3] \citealt{Crockett:2014er} [4] \citealt{Melosso:2018kl} [5] \citealt{Cernicharo:2018bv} [6] \citealt{Higuchi:2018dx} [7] \citealt{McGuire:2015bp} [8]\citealt{Jorgensen:2002ci} [9] \citet{Araki:2017kg} [10] \citealt{HilyBlant:2018ix} [11]  \citealt{Crapsi:2005kp} [12] \citealt{McGuire:2018it} [13] \citealt{Gratier:2016fj} [14] \citealt{Neill:2012fr} [15] \citealt{Belloche:2013eba}
	\end{minipage}    
    \label{h2ccs_source_params}
\end{table*}

\clearpage

\section{Non-Detection Figures}
\label{nondetfigs}

\renewcommand{\thefigure}{B\arabic{figure}}
\renewcommand{\thetable}{B\arabic{table}}
\renewcommand{\theequation}{B\arabic{equation}}
\setcounter{figure}{0}
\setcounter{table}{0}
\setcounter{equation}{0}

\begin{figure*}[!hbt]
    \centering
    \includegraphics[width=0.23\textwidth]{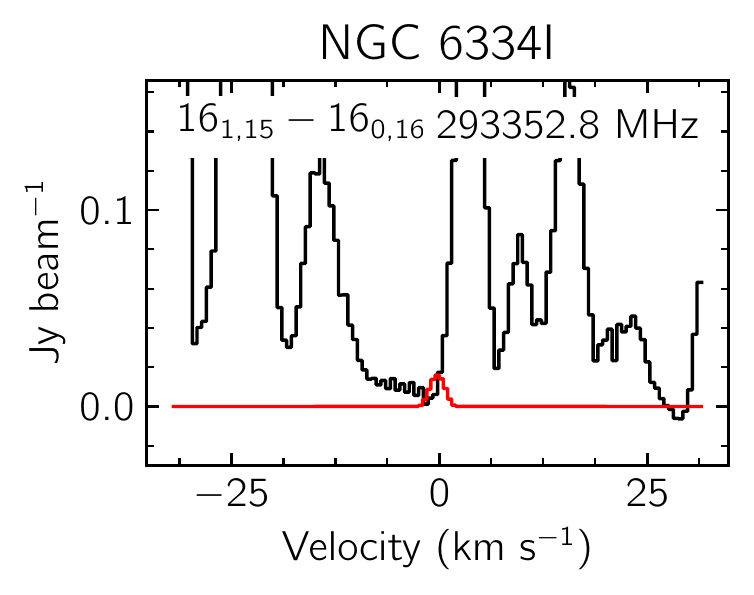}
    \includegraphics[width=0.23\textwidth]{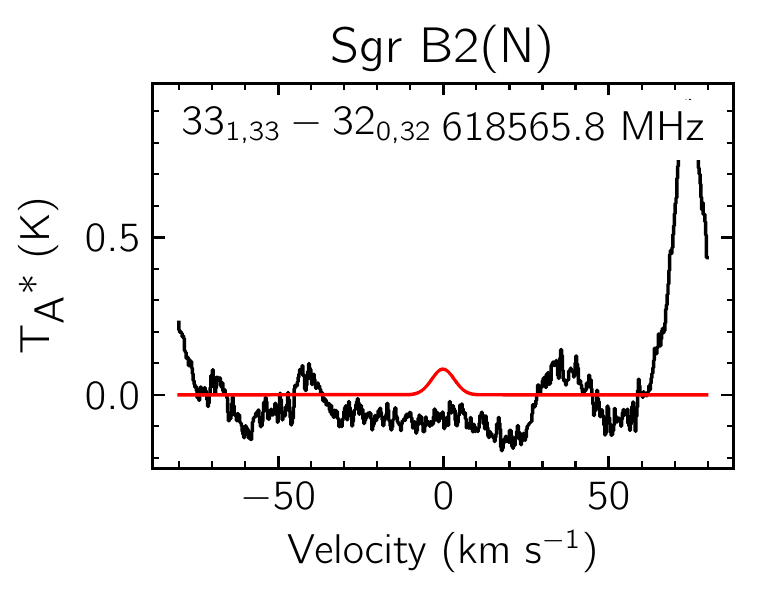}
    \includegraphics[width=0.23\textwidth]{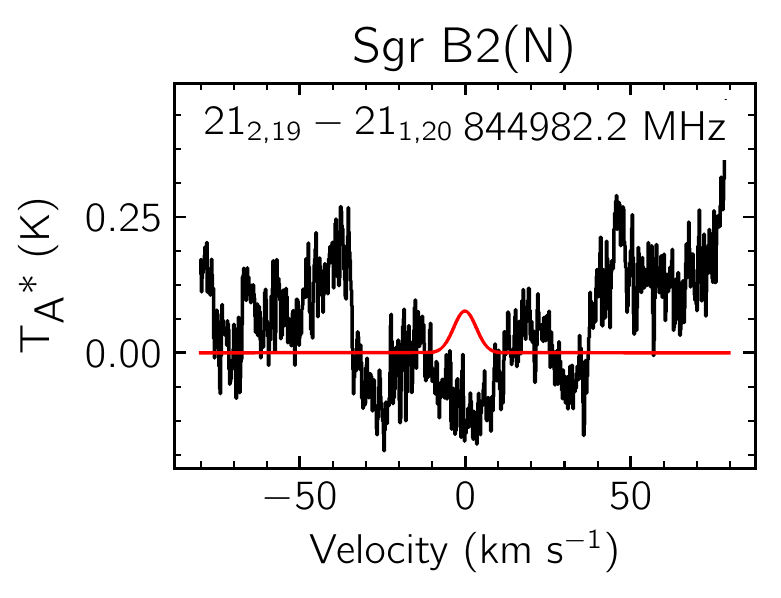}
    \includegraphics[width=0.23\textwidth]{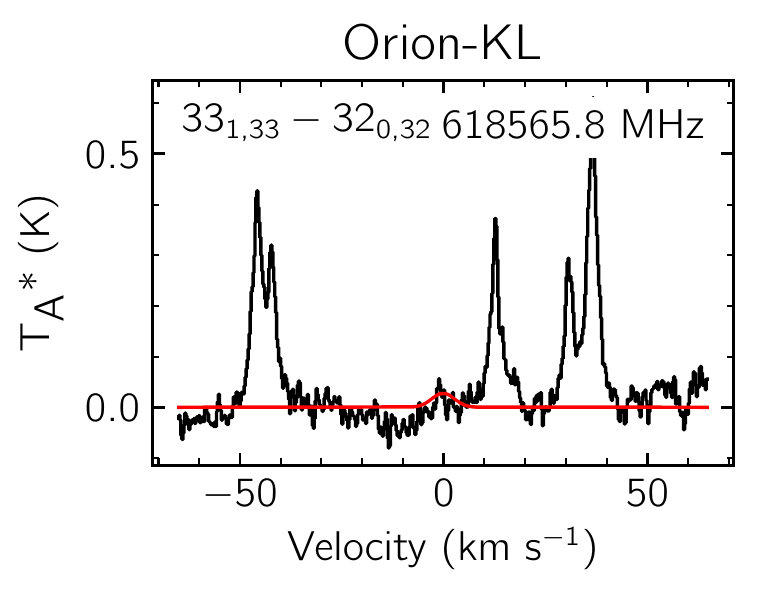}
    \includegraphics[width=0.23\textwidth]{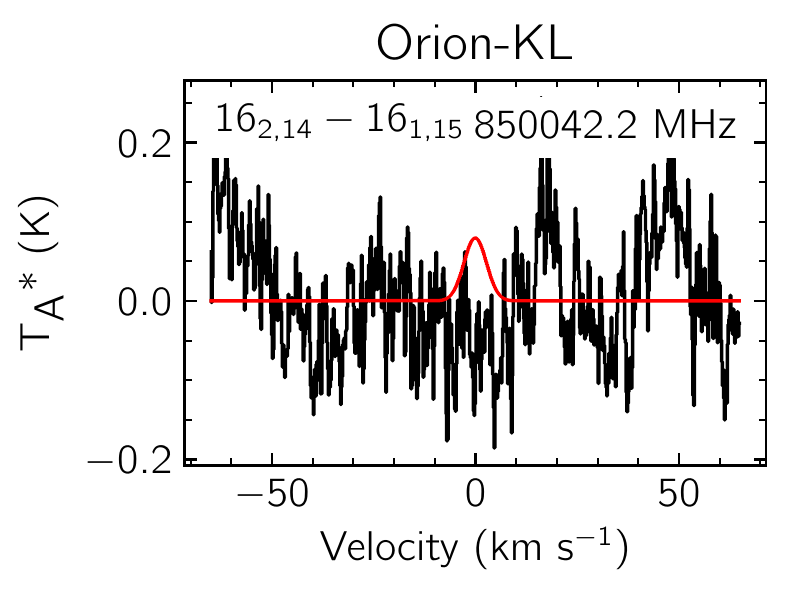}
    \includegraphics[width=0.23\textwidth]{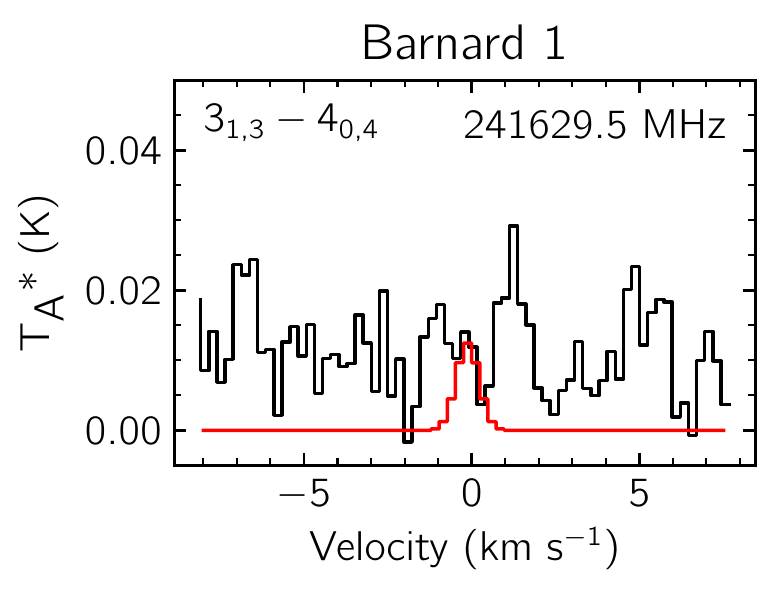}
    \includegraphics[width=0.23\textwidth]{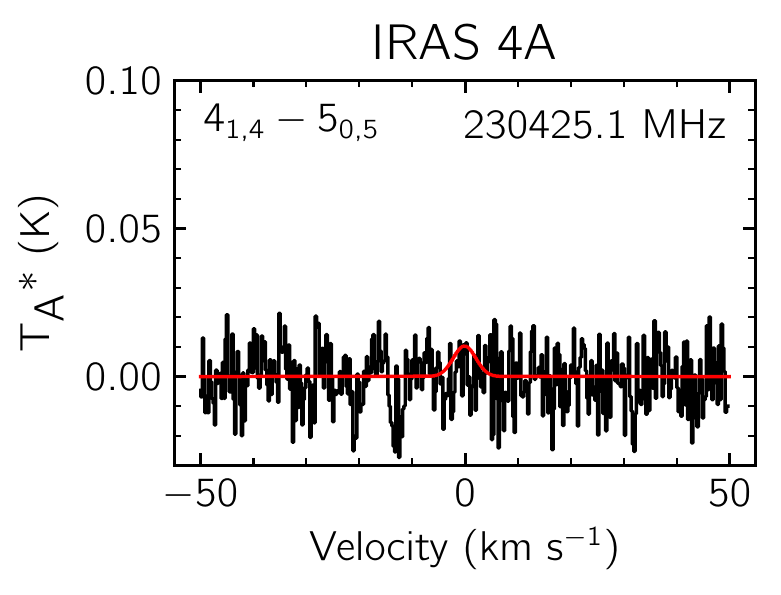}
    \includegraphics[width=0.23\textwidth]{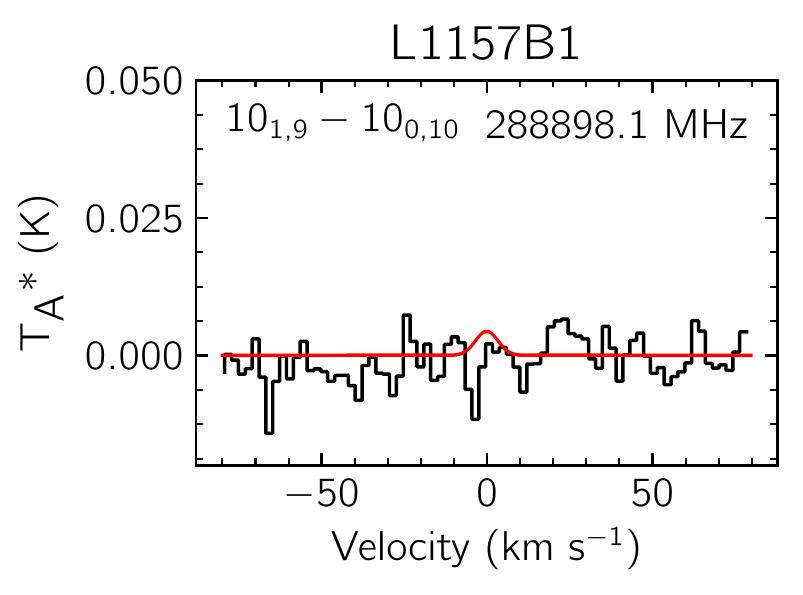}
    \includegraphics[width=0.23\textwidth]{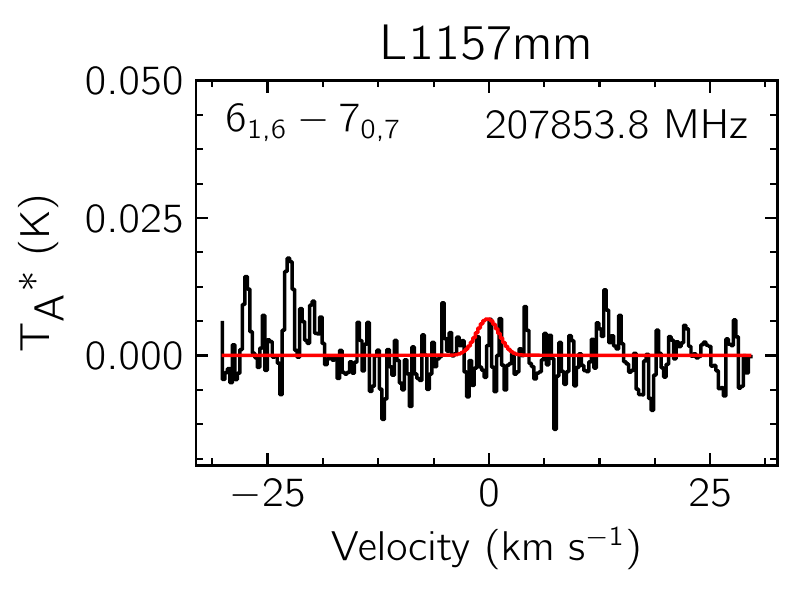}
    \includegraphics[width=0.23\textwidth]{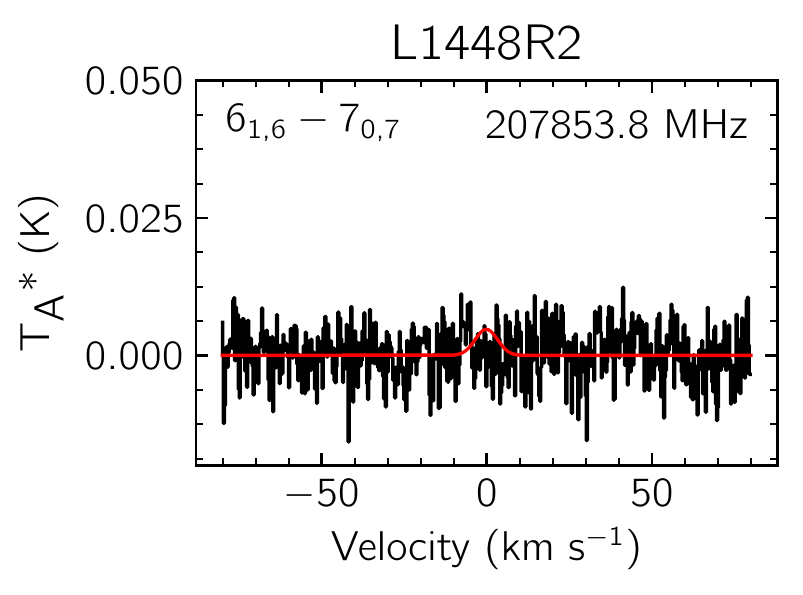}
    \includegraphics[width=0.23\textwidth]{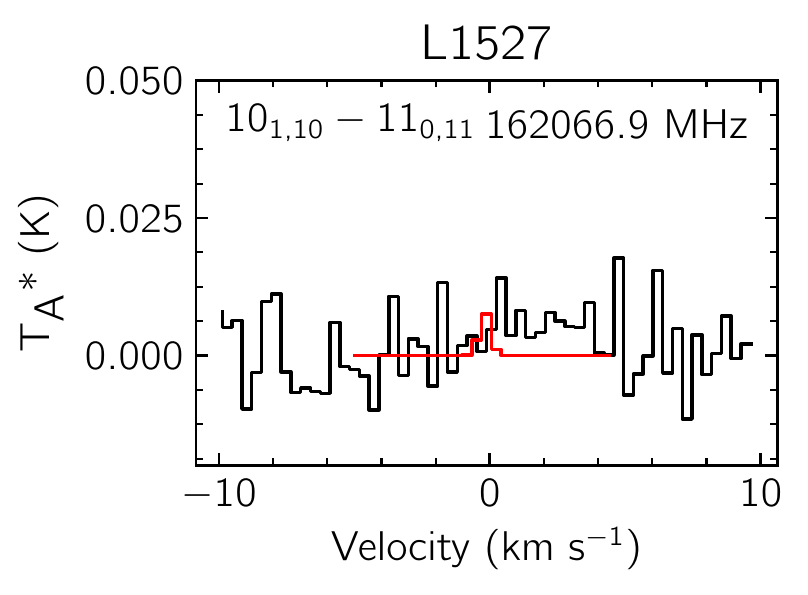}
    \includegraphics[width=0.23\textwidth]{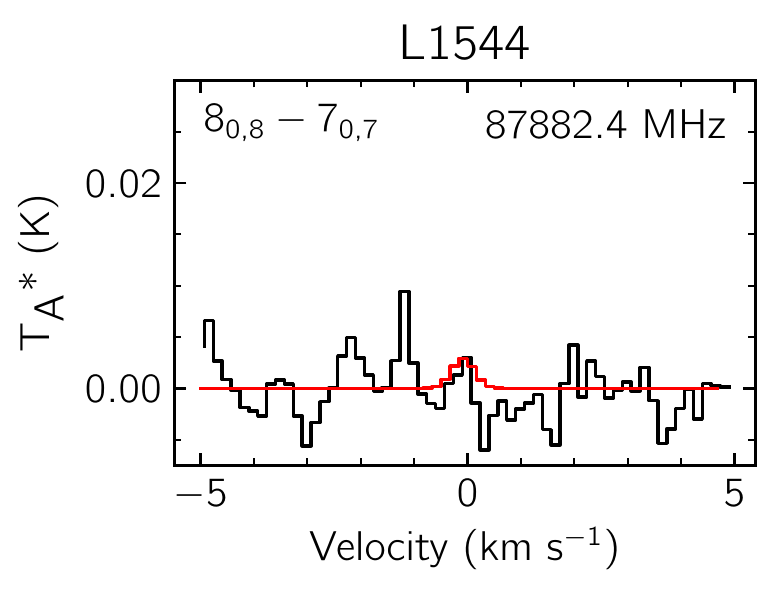}
    \includegraphics[width=0.23\textwidth]{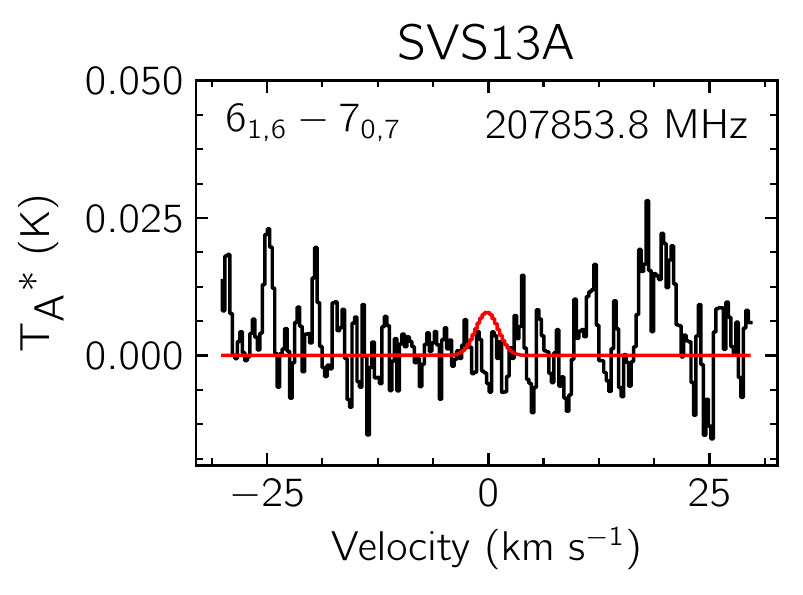}
    \includegraphics[width=0.23\textwidth]{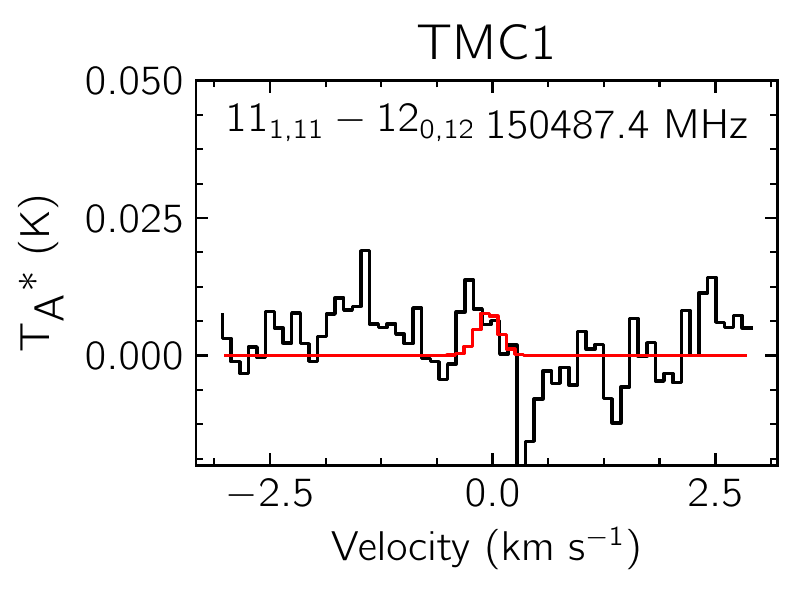}
    \caption{Transitions of HCCSH used to calculate the upper limits given in Table~\ref{hccsh_ulims}.  In each panel, the red trace shows the transition simulated using the derived upper limit column density and the physical parameters assumed for that source.  The frequency of the transition is given in the upper right of each panel, and the quantum numbers for each transition in the upper left.  The source name is given above each panel.  Due to the large variances between observations, the intensity and velocity axes are not uniform between each panel.}
    \label{hccsh_fig}
\end{figure*}

\begin{figure*}[!htb]
    \centering
    \includegraphics[width=0.23\textwidth]{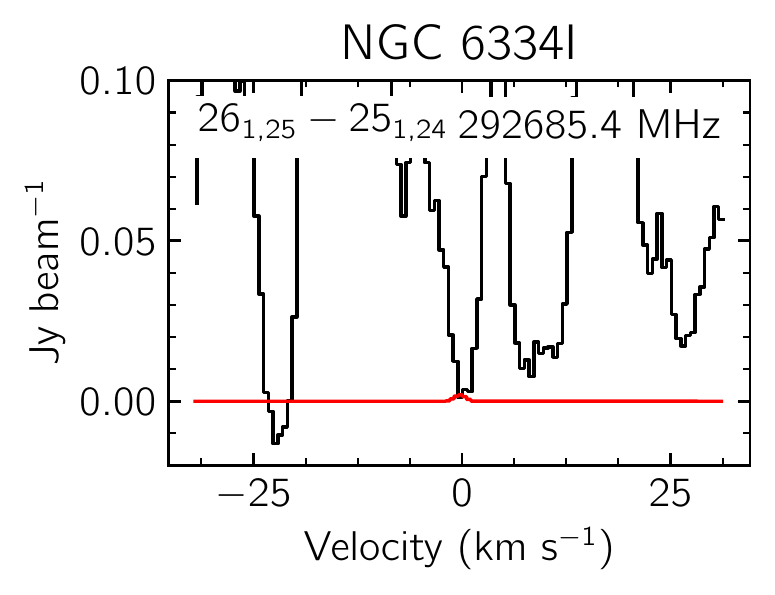}
    \includegraphics[width=0.23\textwidth]{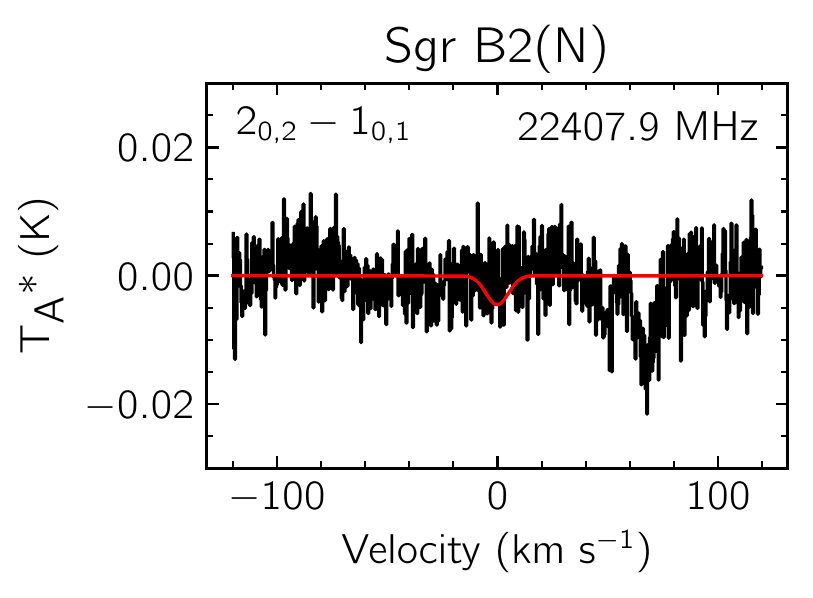}
    \includegraphics[width=0.23\textwidth]{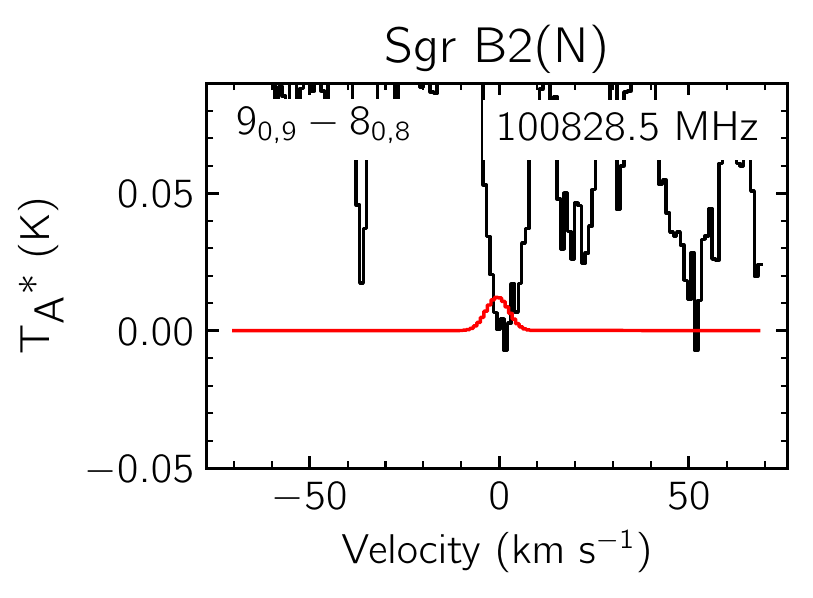}
    \includegraphics[width=0.23\textwidth]{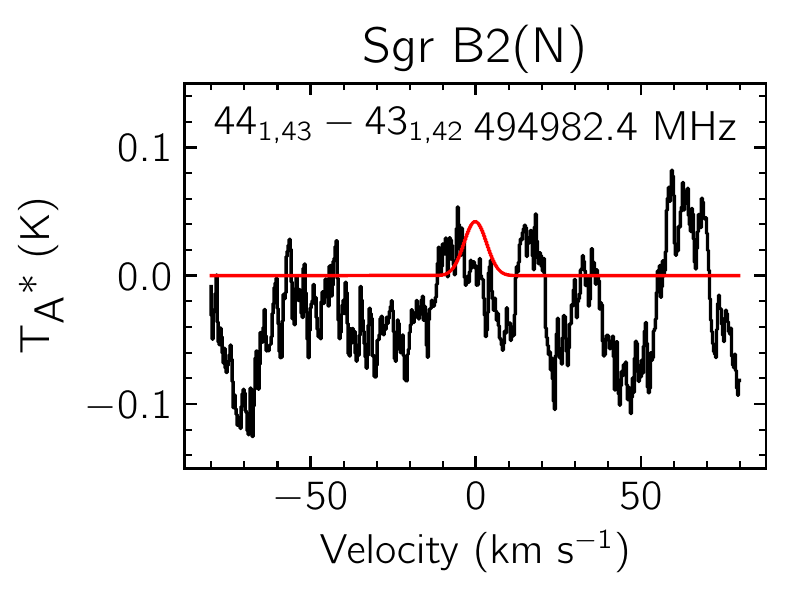}    
    \includegraphics[width=0.23\textwidth]{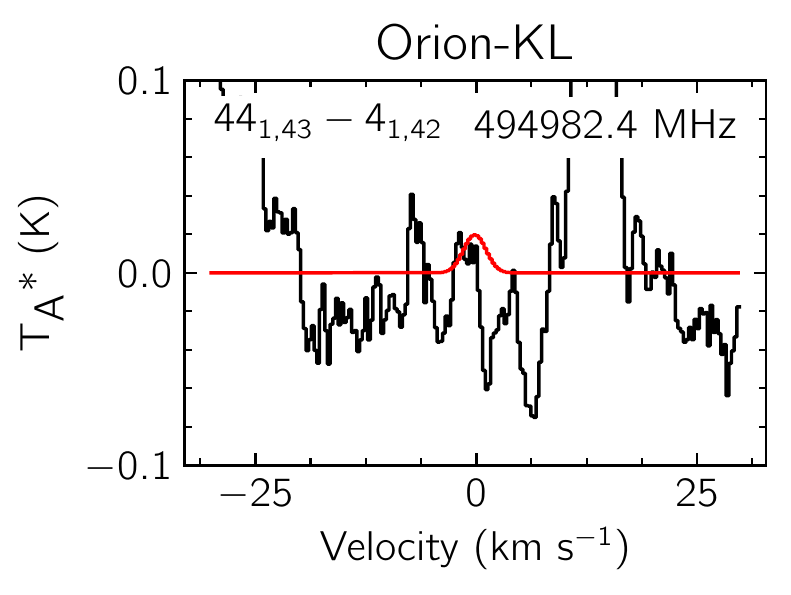}
    \includegraphics[width=0.23\textwidth]{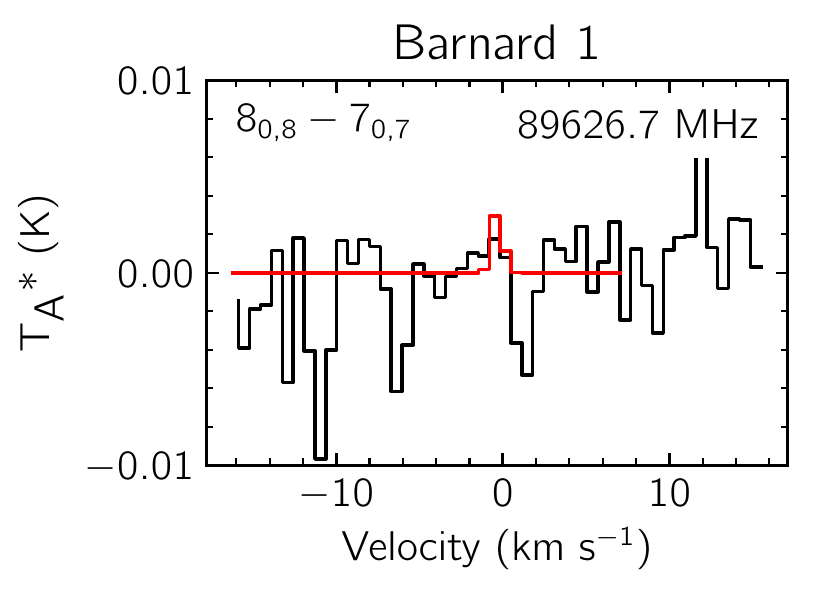}
    \includegraphics[width=0.23\textwidth]{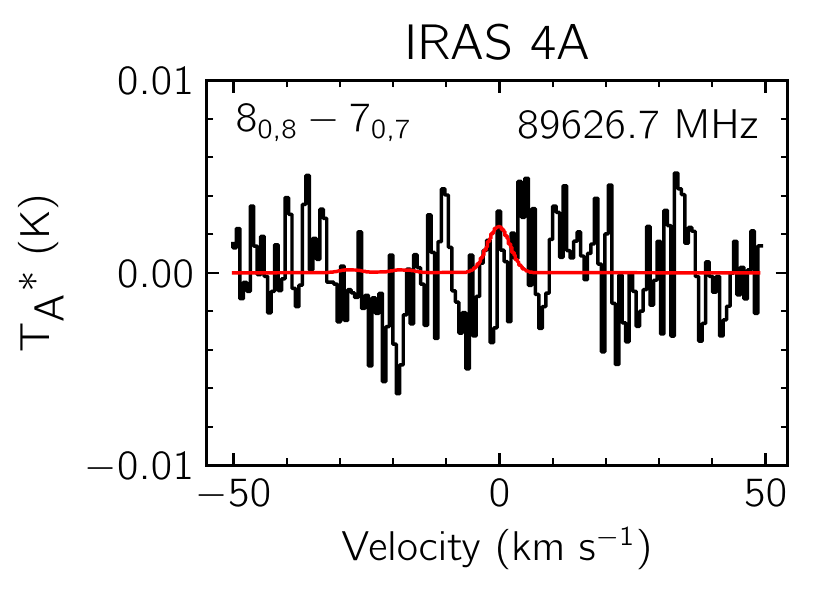}
    \includegraphics[width=0.23\textwidth]{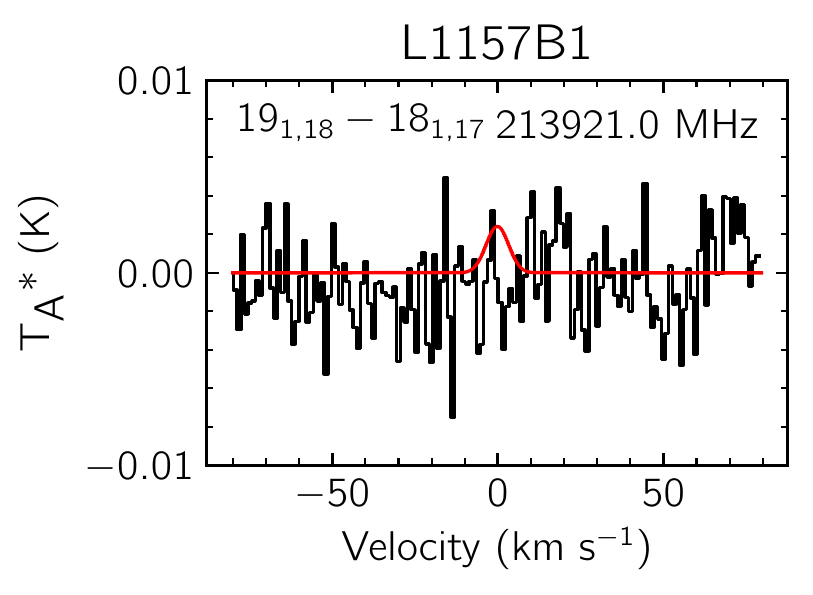}
    \includegraphics[width=0.23\textwidth]{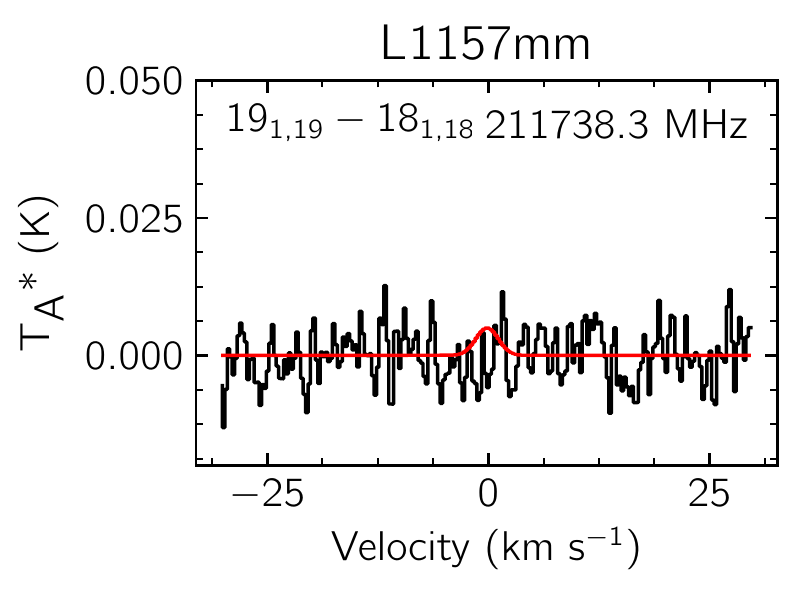}
    \includegraphics[width=0.23\textwidth]{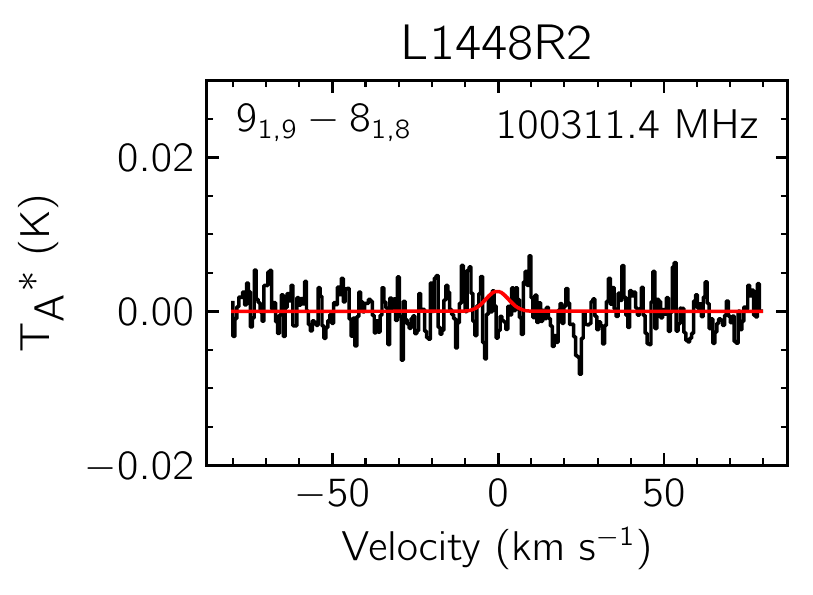}
    \includegraphics[width=0.23\textwidth]{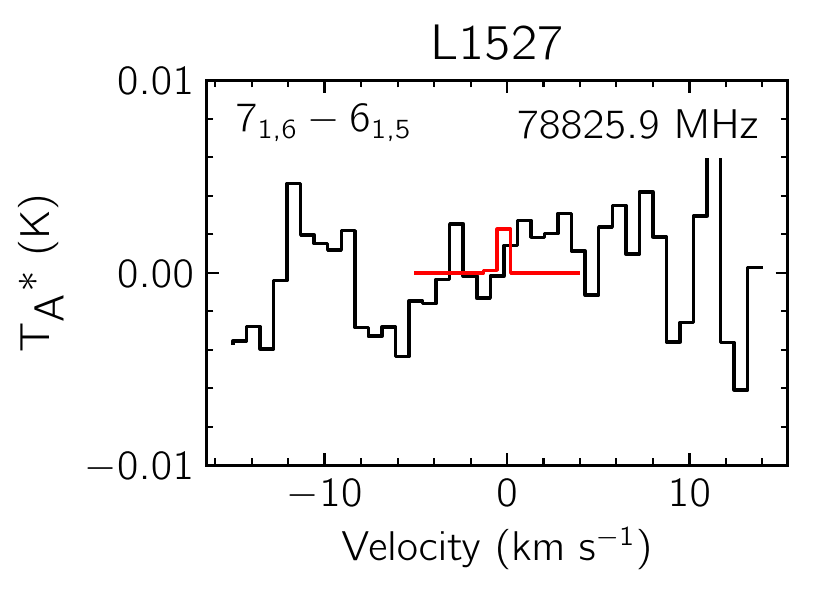}
    \includegraphics[width=0.23\textwidth]{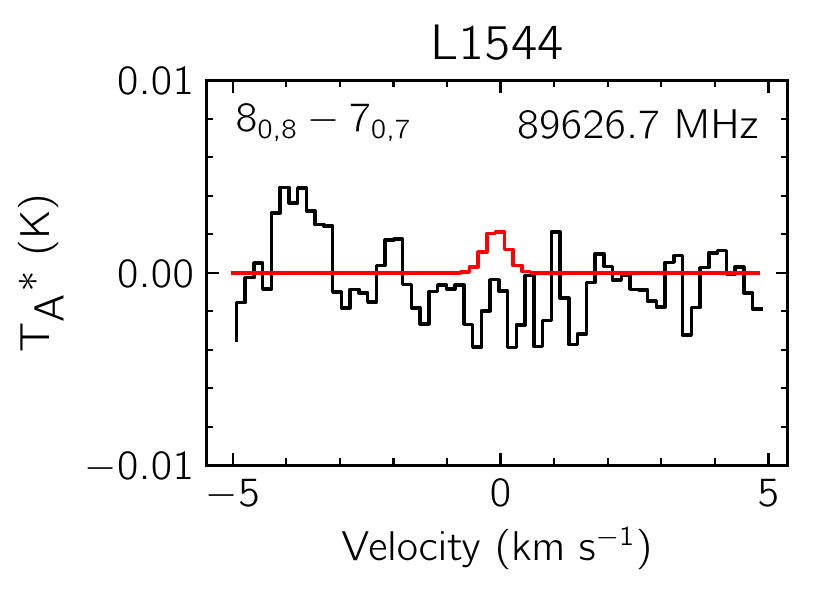}
    \includegraphics[width=0.23\textwidth]{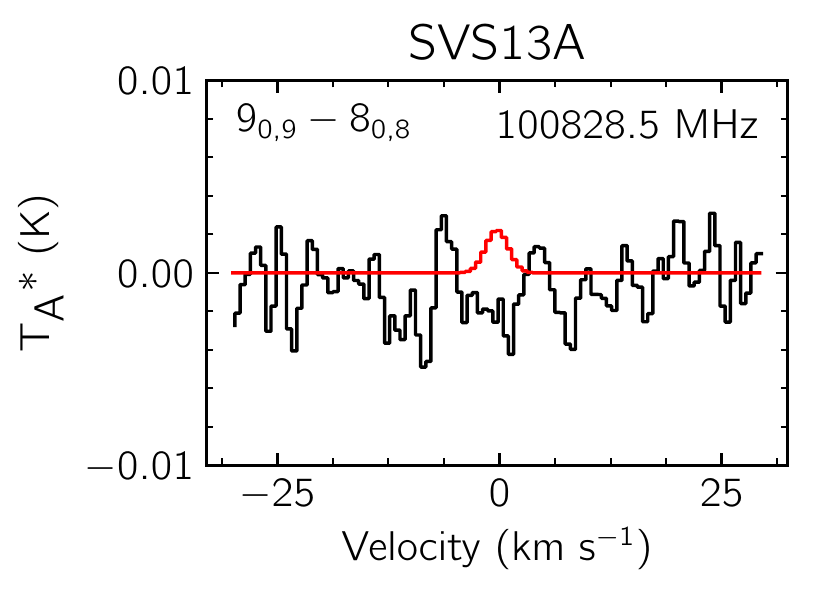}
    \includegraphics[width=0.23\textwidth]{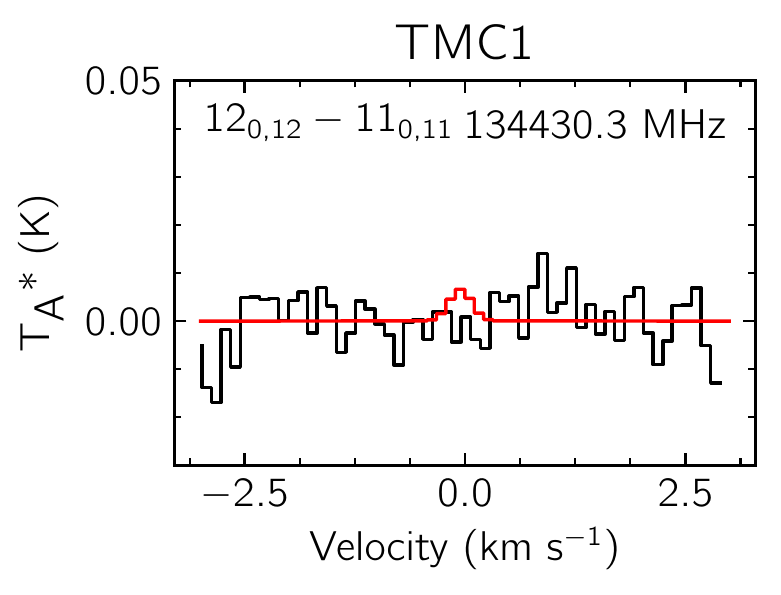}
    \caption{Transitions of \ce{H2CCS} used to calculate the upper limits given in Table~\ref{h2ccs_ulims}.  In each panel, the red trace shows the transition simulated using the derived upper limit column density and the physical parameters assumed for that source.  The frequency of the transition is given in the upper right of each panel, and the quantum numbers for each transition in the upper left.  The source name is given above each panel.  Due to the large variances between observations, the intensity and velocity axes are not uniform between each panel.}
    \label{h2ccs_fig}
\end{figure*}

Figures~\ref{hccsh_fig} and \ref{h2ccs_fig} show the transition used to calculate the upper limit in each source, simulated using the upper limit column density and parameters given in Tables~\ref{hccsh_ulims}, \ref{h2ccs_ulims}, \ref{hccsh_source_params}, and \ref{h2ccs_source_params}.  

\clearpage

\begin{thebibliography}{}
\expandafter\ifx\csname natexlab\endcsname\relax\def\natexlab#1{#1}\fi
\providecommand{\url}[1]{\href{#1}{#1}}

\bibitem[{Ag{\'u}ndez {et~al.}(2018)Ag{\'u}ndez, Marcelino, Cernicharo, \&
  Tafalla}]{Agundez:2018kh}
Ag{\'u}ndez, M., Marcelino, N., Cernicharo, J., \& Tafalla, M. 2018, \aap,
  611, L1

\bibitem[{Araki {et~al.}(2017)Araki, Takano, Sakai, Yamamoto, Oyama, Kuze, \&
  Tsukiyama}]{Araki:2017kg}
Araki, M., Takano, S., Sakai, N., {et~al.} 2017, \apj,
  847, 0

\bibitem[{Bell {et~al.}(1993)Bell, Avery, \& Feldman}]{Bell:1993ir}
Bell, M.~B., Avery, L.~W., \& Feldman, P.~A. 1993, \apj,
  417, L37

\bibitem[{Belloche {et~al.}(2013)Belloche, M{\"u}ller, Menten, Schilke, \&
  Comito}]{Belloche:2013eba}
Belloche, A., M{\"u}ller, H. S.~P., Menten, K.~M., Schilke, P., \& Comito, C.
  2013, \aap, 559, A47

\bibitem[{Benz {et~al.}(2010)Benz, Bruderer, van Dishoeck, St{\"a}uber,
  Wampfler, Melchior, Dedes, Wyrowski, Doty, van~der Tak, B{\"a}chtold,
  Csillaghy, Megej, Monstein, Soldati, Bachiller, Baudry, Benedettini, Bergin,
  Bjerkeli, Blake, Bontemps, Braine, Caselli, Cernicharo, Codella, Daniel,
  di~Giorgio, Dieleman, Dominik, Encrenaz, Fich, Fuente, Giannini, Goicoechea,
  de~Graauw, Helmich, Herczeg, Herpin, Hogerheijde, Jacq, Jellema, Johnstone,
  J{\o}rgensen, Kristensen, Larsson, Lis, Liseau, Marseille, McCoey, Melnick,
  Neufeld, Nisini, Olberg, Ossenkopf, Parise, Pearson, Plume, Risacher,
  Santiago-Garc{\'\i}a, Saraceno, Schieder, Shipman, Stutzki, Tafalla, Tielens,
  van Kempen, Visser, \& Y{\i}ld{\i}z}]{Benz:2010ei}
Benz, A.~O., Bruderer, S., van Dishoeck, E.~F., {et~al.} 2010, \aap, 521, L35

\bibitem[{Bilalbegovi{\'c} \& Baranovi{\'c}(2015)}]{Bilalbegovic:2015dr}
Bilalbegovi{\'c}, G., \& Baranovi{\'c}, G. 2015, \mnras, 446, 3118

\bibitem[{Bockel{\'e}e-Morvan {et~al.}(2000)Bockel{\'e}e-Morvan, Lis, Wink,
  Despois, Crovisier, Bachiller, Benford, Biver, Colom, Davies, G{\'e}rard,
  Germain, Houde, Mehringer, Moreno, Paubert, Phillips, \&
  Rauer}]{Bockelee:2000ol}
Bockel{\'e}e-Morvan, D., Lis, D.~C., Wink, J.~E., {et~al.} 2000, \aap, 353,
  1101

\bibitem[{Bonfand {et~al.}(2017)Bonfand, Belloche, Menten, Garrod, \&
  M{\"u}ller}]{Bonfand:2017eo}
Bonfand, M., Belloche, A., Menten, K.~M., Garrod, R.~T., \& M{\"u}ller, H.
  S.~P. 2017, \aap, 604, A60

\bibitem[{Boogert {et~al.}(2015)Boogert, Gerakines, \&
  Whittet}]{Boogert:2015fx}
Boogert, A. C.~A., Gerakines, P.~A., \& Whittet, D. C.~B. 2015, \araa, 53, 541

\bibitem[{Boogert {et~al.}(1997)Boogert, Schutte, Helmich, Tielens, \&
  Wooden}]{Boogert:1997ys}
Boogert, A. C.~A., Schutte, W.~A., Helmich, F.~P., Tielens, A. G. G.~M., \&
  Wooden, D.~H. 1997, \aap, 317, 929

\bibitem[{Brogan {et~al.}(2016)Brogan, Hunter, Cyganowski, Chandler, Friesen,
  \& Indebetouw}]{Brogan:2016cy}
Brogan, C.~L., Hunter, T.~R., Cyganowski, C.~J., {et~al.} 2016, \apj, 832, 187

\bibitem[{Brogan {et~al.}(2018)Brogan, Hunter, Cyganowski, Chibueze, Friesen,
  Hirota, MacLeod, McGuire, \& Sobolev}]{Brogan:2018wb}
---. 2018, \apj, 866, 87

\bibitem[{Cameron(1973)}]{Cameron:1970dw}
Cameron, A. G.~W. 1973, \ssr, 15, 121

\bibitem[{Cazzoli {et~al.}(2016)Cazzoli, Lattanzi, Kirsch, Gauss, Tercero,
  Cernicharo, \& Puzzarini}]{Cazzoli:2016ew}
Cazzoli, G., Lattanzi, V., Kirsch, T., {et~al.} 2016, \aap, 591, A126

\bibitem[{Cernicharo {et~al.}(1991)Cernicharo, Gottlieb, Gu{\'e}lin, Killian,
  Paubert, Thaddeus, \& Vrtilek}]{Cernicharo:1991iv}
Cernicharo, J., Gottlieb, C.~A., Gu{\'e}lin, M., {et~al.} 1991,  \apj, 368, L39

\bibitem[{Cernicharo {et~al.}(2018)Cernicharo, Lefloch, Ag{\'u}ndez, Bailleux,
  Margul{\`e}s, Roueff, Bachiller, Marcelino, Tercero, Vastel, \&
  Caux}]{Cernicharo:2018bv}
Cernicharo, J., Lefloch, B., Ag{\'u}ndez, M., {et~al.} 2018, \apjl, 853, L22

\bibitem[{Chen {et~al.}(2009)Chen, Launhardt, \& Henning}]{Chen:2009jj}
Chen, X., Launhardt, R., \& Henning, T. 2009, \apj, 691,
  1729

\bibitem[{Chibueze {et~al.}(2014)Chibueze, Omodaka, Handa, Imai, Kurayama,
  Nagayama, Sunada, Nakano, Hirota, \& Honma}]{Chibueze:2014iv}
Chibueze, J.~O., Omodaka, T., Handa, T., {et~al.} 2014, \apj, 784, 114

\bibitem[{Crapsi {et~al.}(2005)Crapsi, Caselli, Walmsley, { P. C. Myers},
  Tafalla, Lee, \& Bourke}]{Crapsi:2005kp}
Crapsi, A., Caselli, P., Walmsley, C.~M., {et~al.} 2005, \apj, 619, 379

\bibitem[{Crockett {et~al.}(2014)Crockett, Bergin, Neill, Favre, Schilke, Lis,
  Bell, Blake, Cernicharo, Emprechtinger, Esplugues, Gupta, Kleshcheva, Lord,
  Marcelino, McGuire, Pearson, Phillips, Plume, van~der Tak, Tercero, \&
  Yu}]{Crockett:2014er}
Crockett, N.~R., Bergin, E.~A., Neill, J.~L., {et~al.} 2014, \apj, 787, 112

\bibitem[{Dai \& Bradley(2001)}]{Dai:2001iu}
Dai, Z.~R., \& Bradley, J.~P. 2001, Geochimica et Cosmochimica Acta, 65, 3601

\bibitem[{Frerking {et~al.}(1979)Frerking, Linke, \&
  Thaddeus}]{Frerking:1979yd}
Frerking, M.~A., Linke, R.~A., \& Thaddeus, P. 1979, \apj,
  234, L143

\bibitem[{Fuente {et~al.}(2017)Fuente, Goicoechea, Pety, Le~Gal,
  Mart{\'\i}n-Dom{\'e}nech, Gratier, Guzm{\'a}n, Roueff, Loison, Caro, Wakelam,
  Gerin, Riviere-Marichalar, \& Vidal}]{Fuente:2017gd}
Fuente, A., Goicoechea, J.~R., Pety, J., {et~al.} 2017, \apjl, 851, L49

\bibitem[{Galland {et~al.}(2001)Galland, Caralp, Rayez, Hannachi, Loison,
  Dorthe, \& Bergeat}]{Galland:2001ei}
Galland, N., Caralp, F., Rayez, M.-T., {et~al.} 2001, The Journal of Physical
  Chemistry A, 105, 9893

\bibitem[{Gordy \& Cook(1984)}]{Gordy:1984uy}
Gordy, W., \& Cook, R.~L. 1984, {Microwave Molecular Spectra}, 3rd edn. (New
  York: Wiley)

\bibitem[{Gottlieb \& Ball(1973)}]{Gottlieb:1973md}
Gottlieb, C.~A., \& Ball, J.~A. 1973, \apj, 184, L59

\bibitem[{Gottlieb {et~al.}(1975)Gottlieb, Ball, Gottlieb, Lada, \&
  Penfield}]{Gottlieb:1975ke}
Gottlieb, C.~A., Ball, J.~A., Gottlieb, E.~W., Lada, C.~J., \& Penfield, H.
  1975, \apj, 200, L147

\bibitem[{Gratier {et~al.}(2016)Gratier, Majumdar, Ohishi, Roueff, Loison,
  Hickson, \& Wakelam}]{Gratier:2016fj}
Gratier, P., Majumdar, L., Ohishi, M., {et~al.} 2016, \apjs, 225, 1

\bibitem[{Halfen {et~al.}(2009)Halfen, Ziurys, Br{\"u}nken, Gottlieb, McCarthy,
  \& Thaddeus}]{Halfen:2009it}
Halfen, D.~T., Ziurys, L.~M., Br{\"u}nken, S., {et~al.} 2009, \apj, 702, L124

\bibitem[{Higuchi {et~al.}(2018)Higuchi, Sakai, Watanabe, L{\'o}pez-Sepulcre,
  Yoshida, Oya, Imai, Zhang, Ceccarelli, Lefloch, Codella, Bachiller, Hirota,
  Sakai, \& Yamamoto}]{Higuchi:2018dx}
Higuchi, A.~E., Sakai, N., Watanabe, Y., {et~al.} 2018, \apjs, 236, 0

\bibitem[{Hily-Blant {et~al.}(2018)Hily-Blant, Faure, Vastel, Magalhaes,
  Lefloch, \& Bachiller}]{HilyBlant:2018ix}
Hily-Blant, P., Faure, A., Vastel, C., {et~al.} 2018, \mnras, 480, 1174

\bibitem[{Holdship {et~al.}(2016)Holdship, Viti, Jimenez-Serra, Lefloch,
  Codella, Podio, Benedettini, Fontani, Bachiller, Tafalla, \&
  Ceccarelli}]{Holdship:2016js}
Holdship, J., Viti, S., Jimenez-Serra, I., {et~al.} 2016, \mnras, 463, 802

\bibitem[{Hollis {et~al.}(2004{\natexlab{a}})Hollis, Jewell, Lovas, \&
  Remijan}]{Hollis:2004uh}
Hollis, J.~M., Jewell, P.~R., Lovas, F.~J., \& Remijan, A. 2004{\natexlab{a}},
  \apj, 613, L45

\bibitem[{Hollis {et~al.}(2004{\natexlab{b}})Hollis, Jewell, Lovas, Remijan, \&
  M{\o}llendal}]{Hollis:2004oc}
Hollis, J.~M., Jewell, P.~R., Lovas, F.~J., Remijan, A., \& M{\o}llendal, H.
  2004{\natexlab{b}}, \apj, 610, L21

\bibitem[{Hollis {et~al.}(2007)Hollis, Jewell, Remijan, \&
  Lovas}]{Hollis:2007ww}
Hollis, J.~M., Jewell, P.~R., Remijan, A.~J., \& Lovas, F.~J. 2007, \apj, 660, L125

\bibitem[{Hollis {et~al.}(2006)Hollis, Remijan, Jewell, \&
  Lovas}]{Hollis:2006ih}
Hollis, J.~M., Remijan, A.~J., Jewell, P.~R., \& Lovas, F.~J. 2006, \apj, 642, 933

\bibitem[{Hunter {et~al.}(2006)Hunter, Brogan, Megeath, Menten, Beuther, \&
  Thorwirth}]{Hunter:2006th}
Hunter, T.~R., Brogan, C.~L., Megeath, S.~T., {et~al.} 2006, \apj, 649, 888

\bibitem[{Hunter {et~al.}(2017)Hunter, Brogan, MacLeod, Cyganowski, Chandler,
  Chibueze, Friesen, Indebetouw, Thesner, \& Young}]{Hunter:2017th}
Hunter, T.~R., Brogan, C.~L., MacLeod, G., {et~al.} 2017, \apjl, 837, L29

\bibitem[{Hunter {et~al.}(2018)Hunter, Brogan, MacLeod, Cyganowski, Chibueze,
  Friesen, Hirota, Smits, Chandler, \& Indebetouw}]{Hunter:2018gz}
Hunter, T.~R., Brogan, C.~L., MacLeod, G.~C., {et~al.} 2018, \apj, 854, 0

\bibitem[{Irvine {et~al.}(1988{\natexlab{a}})Irvine, Brown, Cragg, Friberg,
  Godfrey, Kaifu, Matthews, Ohishi, Suzuki, \& Takeo}]{Irvine:1988dw}
Irvine, W.~M., Brown, R.~D., Cragg, D.~M., {et~al.} 1988{\natexlab{a}}, 
  \apj, 335, L89

\bibitem[{Irvine {et~al.}(1988{\natexlab{b}})Irvine, Friberg, Hjalmarson,
  Ishikawa, Kaifu, Kawaguchi, Madden, Matthews, Ohishi, Saito, Suzuki,
  Thaddeus, Turner, Yamamoto, \& Ziurys}]{Irvine:1988ej}
Irvine, W.~M., Friberg, P., Hjalmarson, {\AA}., {et~al.} 1988{\natexlab{b}},
  \apj, 334, L107

\bibitem[{Jefferts {et~al.}(1971)Jefferts, Penzias, Wilson, \&
  Solomon}]{Jefferts:1971hy}
Jefferts, K.~B., Penzias, A.~A., Wilson, R.~W., \& Solomon, P.~M. 1971,
  \apj, 168, L111

\bibitem[{J{\o}rgensen {et~al.}(2002)J{\o}rgensen, Sch{\"o}ier, \& van
  Dishoeck}]{Jorgensen:2002ci}
J{\o}rgensen, J.~K., Sch{\"o}ier, F.~L., \& van Dishoeck, E.~F. 2002, \aap,
  389, 908

\bibitem[{Keller {et~al.}(2002)Keller, Hony, Bradley, Molster, Waters, Bouwman,
  de~Koter, Brownlee, Flynn, Henning, \& Mutschke}]{Keller:2002db}
Keller, L.~P., Hony, S., Bradley, J.~P., {et~al.} 2002, Nature, 417, 148

\bibitem[{Kolesnikov{\'a} {et~al.}(2014)Kolesnikov{\'a}, Tercero, Cernicharo,
  Alonso, Daly, Gordon, \& Shipman}]{Kolesnikova:2014fb}
Kolesnikov{\'a}, L., Tercero, B., Cernicharo, J., {et~al.} 2014, \apj, 784, L7

\bibitem[{Kuiper {et~al.}(1975)Kuiper, Kakar, Kuiper, \&
  Zuckerman}]{Kuiper:1975er}
Kuiper, T. B.~H., Kakar, R.~K., Kuiper, E. N.~R., \& Zuckerman, B. 1975,
  \apj, 200, L151

\bibitem[{Laas \& Caselli(2019)}]{Laas:2019kb}
Laas, J.~C., \& Caselli, P. 2019, \aap, 624, A108

\bibitem[{Laas {et~al.}(2011)Laas, Garrod, Herbst, \&
  Widicus~Weaver}]{Laas:2011yd}
Laas, J.~C., Garrod, R.~T., Herbst, E., \& Widicus~Weaver, S.~L. 2011, 
  \apj, 728, 71

\bibitem[{Lamberts(2018)}]{Lamberts:2018eh}
Lamberts, T. 2018, \aap, 615, L2

\bibitem[{Lee {et~al.}(2018)Lee, Martin-Drumel, Lattanzi, McGuire, Caselli, \&
  McCarthy}]{Lee:2018uw}
Lee, K. L.~K., Martin-Drumel, M.-A., Lattanzi, V., {et~al.} 2018, Molecular
  Physics, 6, in press

\bibitem[{Lefloch {et~al.}(2018)Lefloch, Bachiller, Ceccarelli, Cernicharo,
  Codella, Fuente, Kahane, Lopez-Sepulcre, Tafalla, Vastel, Caux,
  Gonz{\'a}lez~Garc{\'\i}a, Bianchi, G{\'o}mez-Ruiz, Holdship, Mendoza,
  Ospina-Zamudio, Podio, Quenard, Roueff, Sakai, Viti, Yamamoto, Yoshida,
  Favre, Monfredini, Quiti{\'a}n-Lara, Marcelino, Boechat-Roberty, \&
  Cabrit}]{Lefloch:2018fk}
Lefloch, B., Bachiller, R., Ceccarelli, C., {et~al.} 2018, \mnras, 477, 4792

\bibitem[{Linke {et~al.}(1979)Linke, Frerking, \& Thaddeus}]{Linke:1979dc}
Linke, R.~A., Frerking, M.~A., \& Thaddeus, P. 1979, \apj,
  234, L139

\bibitem[{Lis \& Goldsmith(1990)}]{Lis:1990tw}
Lis, D.~C., \& Goldsmith, P.~F. 1990, \apj, 356, 195

\bibitem[{Loison {et~al.}(2016)Loison, Ag{\'u}ndez, Marcelino, Wakelam,
  Hickson, Cernicharo, Gerin, Roueff, \& Gu{\'e}lin}]{Loison:2016ck}
Loison, J.-C., Ag{\'u}ndez, M., Marcelino, N., {et~al.} 2016, \mnras, 456, 4101

\bibitem[{Loomis {et~al.}(2015)Loomis, McGuire, Shingledecker, Johnson, Blair,
  Robertson, \& Remijan}]{Loomis:2015jh}
Loomis, R.~A., McGuire, B.~A., Shingledecker, C., {et~al.} 2015, 
  \apj, 799, 34

\bibitem[{Mart{\'\i}n-Dom{\'e}nech {et~al.}(2016)Mart{\'\i}n-Dom{\'e}nech,
  Jimenez-Serra, Mu{\~n}oz~Caro, M{\"u}ller, Occhiogrosso, Testi, Woods, \&
  Viti}]{MartinDomenech:2016je}
Mart{\'\i}n-Dom{\'e}nech, R., Jimenez-Serra, I., Mu{\~n}oz~Caro, G.~M.,
  {et~al.} 2016, \aap, 585, A112

\bibitem[{McGuire(2018)}]{McGuire:2018mc}
McGuire, B.~A. 2018, \apjs, 239, 17

\bibitem[{McGuire {et~al.}(2018{\natexlab{a}})McGuire, Burkhardt, Kalenskii,
  Shingledecker, Remijan, Herbst, \& McCarthy}]{McGuire:2018it}
McGuire, B.~A., Burkhardt, A.~M., Kalenskii, S.~V., {et~al.}
  2018{\natexlab{a}}, Science, 359, 202

\bibitem[{McGuire {et~al.}(2016)McGuire, Carroll, Loomis, Finneran, Jewell,
  Remijan, \& Blake}]{McGuire:2016ba}
McGuire, B.~A., Carroll, P.~B., Loomis, R.~A., {et~al.} 2016, Science, 352,
  1449

\bibitem[{McGuire {et~al.}(2015)McGuire, Carroll, Dollhopf, Crockett, Corby,
  Loomis, Burkhardt, Shingledecker, Blake, \& Remijan}]{McGuire:2015bp}
McGuire, B.~A., Carroll, P.~B., Dollhopf, N.~M., {et~al.} 2015, 
  \apj, 812, 1

\bibitem[{McGuire {et~al.}(2017)McGuire, Shingledecker, Willis, Burkhardt,
  El-Abd, Motiyenko, Brogan, Hunter, Margules, Guillemin, Garrod, Herbst, \&
  Remijan}]{McGuire:2017gy}
McGuire, B.~A., Shingledecker, C.~N., Willis, E.~R., {et~al.} 2017, 
  \apjl, 851, L46

\bibitem[{McGuire {et~al.}(2018{\natexlab{b}})McGuire, Brogan, Hunter, Remijan,
  Blake, Burkhardt, Carroll, van Dishoeck, Garrod, Linnartz, Shingledecker, \&
  Willis}]{McGuire:2018bz}
McGuire, B.~A., Brogan, C.~L., Hunter, T.~R., {et~al.} 2018{\natexlab{b}}, 
  \apjl, 863, L35

\bibitem[{Mehringer {et~al.}(1993)Mehringer, Palmer, Goss, \&
  Yusef-Zadeh}]{Mehringer:1993dd}
Mehringer, D.~M., Palmer, P., Goss, W.~M., \& Yusef-Zadeh, F. 1993,
  \apj, 412, 684

\bibitem[{Melosso {et~al.}(2018)Melosso, Melli, Puzzarini, Codella, Spada,
  Dore, Degli~Esposti, Lefloch, Bachiller, Ceccarelli, Cernicharo, \&
  Barone}]{Melosso:2018kl}
Melosso, M., Melli, A., Puzzarini, C., {et~al.} 2018, \aap, 609, A121

\bibitem[{Menten {et~al.}(2007)Menten, Reid, Forbrich, \&
  Brunthaler}]{Menten:2007ew}
Menten, K.~M., Reid, M.~J., Forbrich, J., \& Brunthaler, A. 2007, \aap, 474,
  515

\bibitem[{Morris {et~al.}(1975)Morris, Gilmore, Palmer, Turner, \&
  Zuckerman}]{Morris:1975im}
Morris, M., Gilmore, W., Palmer, P., Turner, B.~E., \& Zuckerman, B. 1975,
  \apj, 199, L47

\bibitem[{Neill {et~al.}(2012)Neill, Muckle, Zaleski, Steber, Pate, Lattanzi,
  Spezzano, McCarthy, \& Remijan}]{Neill:2012fr}
Neill, J.~L., Muckle, M.~T., Zaleski, D.~P., {et~al.} 2012, \apj, 755, 153

\bibitem[{Neill {et~al.}(2014)Neill, Bergin, Lis, Schilke, Crockett, Favre,
  Emprechtinger, Comito, Qin, Anderson, Burkhardt, Chen, Harris, Lord, McGuire,
  McNeill, Monje, Phillips, Steber, Vasyunina, \& Yu}]{Neill:2014cb}
Neill, J.~L., Bergin, E.~A., Lis, D.~C., {et~al.} 2014, \apj, 789, 8

\bibitem[{Neufeld {et~al.}(2012)Neufeld, Falgarone, Gerin, Godard, Herbst,
  Pineau~des For{\^e}ts, Vasyunin, G{\"u}sten, Wiesemeyer, \&
  Ricken}]{Neufeld:2012gz}
Neufeld, D.~A., Falgarone, E., Gerin, M., {et~al.} 2012, \aap, 542, L6

\bibitem[{Ochsenfeld {et~al.}(1999)Ochsenfeld, Kaiser, Lee, \&
  Head-Gordon}]{Ochsenfeld:1999fz}
Ochsenfeld, C., Kaiser, R.~I., Lee, Y.~T., \& Head-Gordon, M. 1999, The Journal
  of Chemical Physics, 110, 9982

\bibitem[{Penzias {et~al.}(1971)Penzias, Solomon, Wilson, \&
  Jefferts}]{Penzias:1971kw}
Penzias, A.~A., Solomon, P.~M., Wilson, R.~W., \& Jefferts, K.~B. 1971,
  \apj, 168, L53

\bibitem[{Ray(1932)}]{Ray:1932yd}
Ray, B.~S. 1932, Zeitschrift f{\"u}r Physik, 78, 74

\bibitem[{Reid {et~al.}(2014)Reid, Menten, Brunthaler, Zheng, Dame, Xu, Wu,
  Zhang, Sanna, Sato, Hachisuka, Choi, Immer, Moscadelli, Rygl, \&
  Bartkiewicz}]{Reid:2014km}
Reid, M.~J., Menten, K.~M., Brunthaler, A., {et~al.} 2014, \apj, 783, 130

\bibitem[{Ruffle {et~al.}(1999)Ruffle, Hartquist, Caselli, \&
  Williams}]{Ruffle:1999fla}
Ruffle, D.~P., Hartquist, T.~W., Caselli, P., \& Williams, D.~A. 1999, \mnras, 306, 691

\bibitem[{Saito {et~al.}(1987)Saito, Kawaguchi, Yamamoto, Ohishi, Suzuki, \&
  Kaifu}]{Saito:1987fa}
Saito, S., Kawaguchi, K., Yamamoto, S., {et~al.} 1987, \apj,
  317, L115

\bibitem[{Shingledecker {et~al.}(2019)Shingledecker, {\'A}lvarez-Barcia, Korn,
  \& K{\"a}stner}]{Shingledecker:2019js}
Shingledecker, C.~N., {\'A}lvarez-Barcia, S., Korn, V.~H., \& K{\"a}stner, J.
  2019, \apj, 878, 0

\bibitem[{Sinclair {et~al.}(1973)Sinclair, Fourikis, Ribes, Robinson, Brown, \&
  Godfrey}]{Sinclair:1973ft}
Sinclair, M.~W., Fourikis, N., Ribes, J.~C., {et~al.} 1973, Australian Journal
  of Physics, 26, 85

\bibitem[{Snyder {et~al.}(1975)Snyder, Hollis, Ulich, Lovas, Johnson, \&
  Buhl}]{Snyder:1975cr}
Snyder, L.~E., Hollis, J.~M., Ulich, B.~L., {et~al.} 1975, \apj, 198, L81

\bibitem[{Thaddeus {et~al.}(1981)Thaddeus, Gu{\'e}lin, \&
  Linke}]{Thaddeus:1981uy}
Thaddeus, P., Gu{\'e}lin, M., \& Linke, R.~A. 1981, \apj, 246,
  L41

\bibitem[{Thaddeus {et~al.}(1972)Thaddeus, Kutner, Penzias, Wilson, \&
  Jefferts}]{Thaddeus:1972oh}
Thaddeus, P., Kutner, M.~L., Penzias, A.~A., Wilson, R.~W., \& Jefferts, K.~B.
  1972, \apj, 176, L73

\bibitem[{Tieftrunk {et~al.}(1994)Tieftrunk, Pineau~des For{\^e}ts, Schilke, \&
  Walmsley}]{Tieftrunk:1994rk}
Tieftrunk, A., Pineau~des For{\^e}ts, G., Schilke, P., \& Walmsley, C.~M. 1994,
  \aap, 289, 579

\bibitem[{Turner(1977)}]{Turner:1977ig}
Turner, B.~E. 1977, \apj, 213, L75

\bibitem[{Turner(1991)}]{Turner:1991um}
---. 1991, \apjs, 76, 617

\bibitem[{Turner(1992)}]{Turner:1992cj}
---. 1992, \apj, 396, L107

\bibitem[{Vastel {et~al.}(2014)Vastel, Ceccarelli, Lefloch, \&
  Bachiller}]{Vastel:2014ev}
Vastel, C., Ceccarelli, C., Lefloch, B., \& Bachiller, R. 2014, 
  \apjl, 795, L2

\bibitem[{Vidal {et~al.}(2017)Vidal, Loison, Jaziri, Ruaud, Gratier, \&
  Wakelam}]{Vidal:2017gwa}
Vidal, T. H.~G., Loison, J.-C., Jaziri, A.~Y., {et~al.} 2017, \mnras, 469, 435

\bibitem[{Winnewisser \& Sch{\"a}fer(1980)}]{Winnewisser:1980ic}
Winnewisser, M., \& Sch{\"a}fer, E. 1980, Zeitschrift f{\"u}r Naturforschung A,
  35

\bibitem[{Yamada {et~al.}(2002)Yamada, Osamura, \& Kaiser}]{Yamada:2002gm}
Yamada, M., Osamura, Y., \& Kaiser, R.~I. 2002, \aap, 395, 1031

\bibitem[{Yamamoto {et~al.}(1987)Yamamoto, Saito, Kawaguchi, Kaifu, Suzuki, \&
  Ohishi}]{Yamamoto:1987jd}
Yamamoto, S., Saito, S., Kawaguchi, K., {et~al.} 1987, \apj,
  317, L119

\end{thebibliography}
\end{document}